\documentclass[prb, twocolumn, showpacs,10pt,superscriptaddress]{revtex4-1}
\pdfoutput=1

\usepackage{amssymb,amsfonts,amsmath} 
\usepackage{graphicx}
\usepackage{hyperref}
\usepackage{color}
\usepackage{graphics}
\usepackage{physics}
\usepackage{tikz}
\usepackage{dsfont}
\usepackage{enumitem}
\usepackage{appendix}

\bibpunct{[}{]}{,}{n}{}{}

\definecolor{lightblue}{RGB}{89, 171, 227}

\definecolor{change}{RGB}{255, 0, 0}

\DeclareMathOperator{\sgn}{sgn}

\begin{document} 

\title{
Finite-bias transport through the interacting resonant level model coupled to a phonon mode -- a functional renormalization group study
}

\author{M.~Caltapanides}
\affiliation{Institut f{\"u}r Theorie der Statistischen Physik, RWTH Aachen University and
  JARA---Fundamentals of Future Information Technology, 52056 Aachen, Germany}
\author{D.~M.~Kennes}
\affiliation{Institut f{\"u}r Theorie der Statistischen Physik, RWTH Aachen University and
  JARA---Fundamentals of Future Information Technology, 52056 Aachen, Germany}
 \affiliation{Max Planck Institute for the Structure and Dynamics of Matter, Center for Free Electron Laser Science, 22761 Hamburg, Germany}
\author{V.~Meden}
\affiliation{Institut f{\"u}r Theorie der Statistischen Physik, RWTH Aachen University and
  JARA---Fundamentals of Future Information Technology, 52056 Aachen, Germany}

\begin{abstract}
We study the nonlinear steady-state transport of spinless fermions through a quantum dot with a local two-particle interaction. The dot degree of freedom is in addition coupled to a phonon mode. This setup combines the nonequilibrium physics of the interacting resonant level model and that of the Anderson-Holstein model. The fermion-fermion interaction defies a perturbative treatment. We mainly focus on the antiadiabatic limit, with the phonon frequency being larger than the lead-dot tunneling rate. In this regime also the fermion-boson coupling cannot be treated perturbatively. Our goal is two-fold. We investigate the competing roles of the fermion-fermion and fermion-boson interactions on the emergent low-energy scale $T_{\rm K}$ and show how $T_{\rm K}$ manifests in the transport coefficients as well as the current-voltage characteristics. For small to intermediate interactions, the latter is in addition directly affected by both interactions independently. With increasing fermion-boson interaction the Franck-Condon blockade suppresses the current at small voltages and the emission of phonons leads to shoulders or steps at multiples of the phonon frequency, while the local fermion-fermion interaction implies a negative differential conductance at voltages larger than $T_{\rm K}$. We, in addition, use the model to investigate the limitations of our low-order truncated functional renormalization group approach on the Keldysh contour. In particular, we quantify the role of the broken current conservation. 
\end{abstract}

\date{\today} 
\maketitle

\section{Introduction}
\label{sec:intro}

Emergent many-body phenomena are a hallmark of bulk correlated quantum materials \cite{Giustin2021}. However, one of the most famous of such collective behavior, the Kondo effect \cite{Hewson1993}, occurs in systems with only a few interacting degrees of freedom. Although the Kondo effect was first observed in bulk materials contaminated by magnetic (quantum) impurities, later on mesoscopic systems allowed for rather detailed studies \cite{Pustilnik2004,Grobis2006}.

Mesoscopic devices coupled to leads provide a controlled and tunable environment to investigate locally correlated systems. One of the simplest examples of this type is a quantum dot with spin-degenerate single-particle levels and level spacings, which are the largest energy scale of the problem. In such effective single-level dots the screened, local Coulomb interaction cannot be neglected, in particular, at low energies. If the dot energy is properly tuned such that a single electron occupies the dot, the magnetic exchange interaction between the dot's spin-1/2 degree of freedom and the spins of the lead electrons as well as the associated spin fluctuations will lead to the Kondo effect. The realization of the Kondo effect in mesoscopic transport geometries added another twist to the problem, namely, the steady-state nonequilibrium resulting out of a finite bias voltage applied across the dot. The Kondo effect under finite bias poses a long-standing open problem with certain aspects remaining unclear to this day (see, e.g., Ref.~\onlinecite{Reininghaus2017} and references therein). 

A less known but equally fascinating emergent many-body phenomenon is found in a spinless (spin-polarized) quantum dot setup. In its simplest version the fermion occupying a single impurity level, is coupled via a repulsive, screened, local Coulomb interaction $U$ to the fermions located at the boundaries of two reservoirs. The impurity level is broadened by tunnel couplings $\Gamma_{L/R}$ to the reservoirs. The level energy $\epsilon$ can be tuned by an applied gate voltage. The corresponding model is known as the interacting resonant level model (IRLM) \cite{Wiegmann1978,Schlottmann1980,Filyov1980,Schlottmann1982,Tsvelick1983}. In this setup correlated charge fluctuations prevail at low temperatures. In analogy to the Kondo effect, they lead to an emergent low energy scale which, in reminiscence of the latter, is denoted as $T_{\rm K}$. As in Kondo systems, observables become universal functions of energy variables rescaled by $T_{\rm K}$ \cite{Tsvelick1983}. In the IRLM $T_{\rm K}$ is a measure for the lead-dot tunnel coupling $\Gamma=\Gamma_L+\Gamma_R$ renormalized by the two-particle interaction $U$. For an energy $\epsilon$, such that the dot is half filled (particle-hole symmetric point), lowest-order perturbation theory in $U$ reveals a correction to $\Gamma$ of the form $-U \ln(\Gamma/D)$, where $D$ is a measure for the reservoir band width. This indicates that in the limit of a well-defined dot $\Gamma/D \ll 1$ and for a repulsive two-particle interaction $U>0$ the tunneling rate increases. However, the logarithmic increase for $\Gamma/D \to 0$ shows that perturbative (in either $U$ or $\Gamma$) methods fail in this so-called scaling limit. Several approaches to avoid this, being either analytical \cite{Schlottmann1980,Filyov1980,Schlottmann1982,Tsvelick1983,Borda2007,Borda2008,Karrasch2010} or numerical \cite{Bohr2007,Borda2007,Borda2008,Karrasch2010,Nghiem2016} are available.

It is well established \cite{Schlottmann1980,Filyov1980,Schlottmann1982,Tsvelick1983,Borda2007,Borda2008,Karrasch2010} that the above logarithmic term is the first of a series of leading logarithms of the form $U^n \ln^n(\Gamma/D)$, $n \in {\mathbb N}$. After resummation one obtains
\begin{align} 
	\frac{T_{\rm K}}{\Gamma} = \left( \frac{\Gamma}{D}\right)^{-\alpha_{\Gamma}(U)} , 
	\label{eq:IRLMscale}
\end{align}
with a $U$-dependent exponent $\alpha_{\Gamma}$. For small to intermediate positive values of $U$, $\alpha_{\Gamma}>0$. 

Several of the many-body methods referred to above can also be used to tackle the IRLM in the bias-voltage $V$ or temperature gradient ($T_L \neq T_R$, with the reservoir temperatures $T_{L/R}$) driven nonequilibrium steady state \cite{Doyon2007,Boulat2007,Borda2007,Karrasch2010,Karrasch2010b,Andergassen2011,Kennes2013,Kennes2013b,Freton2014,Culver2020}. The physics of the IRLM is thus well understood also in this regime, in particular, for the case of a left-right symmetric setup ($\Gamma_L = \Gamma_R$, $U_L = U_R$) on which we focus. Throughout this paper we assume that the bias voltage is applied symmetrically to the leads, $V/2$ on the left and $-V/2$ on the right with $V \geq 0$. The main effect of the interplay between the local correlations and a finite bias is a negative differential conductance for $ T_{\rm K} \ll V \ll D$ [equilibrium $T_{\rm K}$; see Eq.~(\ref{eq:IRLMscale})] with the (particle) current $I^{\rm N}$ following the power law suppression 
\begin{equation}
	I^{\rm N} \sim (V/T_{\rm K})^{-\alpha_{I}(U)}
	\label{eq:currIRLM}
\end{equation}
for increasing $V$ \cite{Doyon2007,Boulat2007,Karrasch2010,Kennes2013}. To leading order in $U$ one finds $\alpha_\Gamma = \alpha_{I}$. More involved behavior is found if the left-right symmetry is broken \cite{Karrasch2010b,Kennes2013}.    

Another spinless locally correlated model of interest is one in which the occupation of the dot level can lead to the emission or absorption of single-mode phonons with amplitude $\lambda$. If the dot is realized by a molecule instead of, e.g., a seminconductor heterostructure, this phonon mode represents the dominant vibrational degree of freedom \cite{Mitra2004,Koch2005,Leturcq2009}. The model is known as the spinless Anderson-Holstein model (SAHM). Also this model is well studied in equilibrium as well as in steady-state nonequilibrium \cite{Koch2011,Huetzen2012,Eidelstein2013,Jovchev2013,Khedri2017,Khedri2017b,Khedri2018}.  

With the phonon frequency $\omega_0$ an additional energy scale enters the problem. For phonons which are slow compared to the fermion tunneling $\omega_0 \ll \Gamma$ (adiabatic regime), the physics is barely affected by the phonons and the fermion-boson interaction can be treated in low-order perturbation theory \cite{Mitra2004}. However, for phonons sufficiently fast to react to the fermion tunneling $\Gamma \ll \omega_0$ (antiadiabatic regime), interesting many-body effects can be found \cite{Koch2011,Huetzen2012,Eidelstein2013,Jovchev2013,Khedri2017,Khedri2017b,Khedri2018}. Therefore, we focus on this regime.

Integrating out the phonons \cite{Negele1998} leads to a model with a retarded but purely local and attractive fermion-fermion interaction of strength $\sim \lambda^2$ of the dot degree of freedom; see, e.g., Ref.~\onlinecite{Laakso2014}. This has to be contrasted to the repulsive, instantaneous nearest-neighbor interaction of the IRLM. 

The fermion-phonon interaction leads to a renormalization of the tunnel-coupling similar to that found in the IRLM, however, in the opposite direction \cite{Eidelstein2013,Khedri2017}. While in the $U>0$ IRLM $T_{\rm K} > \Gamma$ for the SAHM one finds $T_{\rm K} < \Gamma$. In the antiadiabatic limit the latter renormalization of the low-energy scale is again logarithmic in the bare tunnel coupling $\Gamma$ with a prefactor $\lambda^2$. Using an RG method by which the leading logarithms are resummed one obtains \cite{Khedri2017}
\begin{align}
	\frac{T_{\rm K}}{\Gamma} = e^{ - \left( \frac{\lambda}{\omega_0}\right)^2}
	\left( \frac{\Gamma}{\omega_0}\right)^{-\frac{4 \Gamma}{\pi \omega_0}\left( \frac{\lambda}{\omega_0} \right)^2} ,
	\label{eq:SAHMscale}
\end{align}
which (only) holds for $\Gamma/\omega_0 \ll 1$ and $\lambda/\omega_0 \lessapprox 1$ only. This, however, is the regime of boson frequencies and fermion-boson couplings of interest to us. For $\Gamma/\omega_0 \ll 1$ (antiadiabatic limit) but all $\lambda$ an alternative expression for $T_{\rm K}$ can be derived by mapping the SAHM to the IRLM and employing the result for $T_{\rm K}$ for the latter. For small $\lambda$, it agrees with Eq.~(\ref{eq:SAHMscale}) \cite{Eidelstein2013}. This mapping should not be confused with the integrating out of the phonons mentioned above, which can be employed for all $\Gamma/\omega_0$ but does not lead to an effective IRLM. The first factor of Eq.~(\ref{eq:SAHMscale}) is the exponential polaronic reduction of the tunneling rate \cite{Lang1963}. The second one is of IRLM-type.   

Also the $I^{\rm N}(V)$ characteristics of the SAHM model is well studied and understood in detail. Away from half filling of the dot, the phonons lead to a Franck-Condon blockade \cite{Koch2005,Leturcq2009}. For increasing $\lambda$ the current at small voltages decreases \cite{Koch2011,Huetzen2012,Khedri2018}. In addition, in the antiadiabatic regime the phonons induce shoulders in $I^{\rm N}$ located at voltages $V \approx 2 \epsilon + 2 n  \omega_0$, $n \in {\mathbb N}$. These result from phonon satellites in the single-particle spectral function which enter the transport energy window when increasing the voltage \cite{Laakso2014,Khedri2018}. For small to intermediate $\lambda$ of interest to us, only the first one or two of these shoulders are visible. For large voltages the current approaches a $\Gamma/\omega_0$-dependent constant value which is, however, independent of $\lambda$. The power-law suppression of the current (negative differential conductance), as discussed for the IRLM, is not found. 

The goal of the present work is two-fold. Using an approximate, truncated functional renormalization group (RG) approach on the Keldysh contour, which proved its power for both the IRLM and the SAHM separately, we investigate the interplay of the local Coulomb interaction and the local fermion-boson coupling. The corresponding model we refer to as the combined model (CM). We show how a low-energy scale $T_{\rm K}$ obtained by combining Eqs.~(\ref{eq:IRLMscale}) and (\ref{eq:SAHMscale}) manifests in the observables. The current and the transport coefficients are computed and show very rich behavior. 

We furthermore use the present model to investigate the limitations of truncated functional RG. We consider the scheme in which one derives an infinite hierarchy of coupled differential flow equations for the one-particle irreducible vertex functions \cite{Metzner2012,Kopietz2010}. In the standard truncation procedure to order $m_{\rm c}$, one sets the vertex functions of order $m > m_{\rm c}$ to their initial values, the bare $m$-particle interactions (being zero for $m>2$ if bare two-particle interactions are considered). This way one obtains an approximation for the vertices which contain all diagrams up to order $m_{\rm c}$. However, by the RG procedure, certain classes of higher order diagrams are captured as well. It was, for example, shown that using this procedure to order $m_{\rm c}=1$ for the IRLM, the series of leading logarithms for $T_{\rm K}$ is fully captured, the power-law suppression of the current at large voltages properly described, and $\alpha_\Gamma$ as well as $\alpha_{I}$ obtained by this approximation are correct to leading order in $U$ \cite{Karrasch2010,Kennes2013}. The situation can be analyzed in even more detail for the somewhat simpler, but related, x-ray edge singularity problem \cite{Diekmann2020}. An application of the functional RG truncated at lowest order to the SAHM revealed all the many-body features of this model mentioned above \cite{Khedri2017,Khedri2017b,Khedri2018}.

There is, however, a drawback of the standard truncation procedure. It does not constitute a conserving approximation \cite{Enss2005}. Fundamental conservation laws, such as, e.g., particle conservation, and thus current conservation, are only guaranteed to be fulfilled up to order $m_{\rm c}$ in the interaction. For models in which the (bare) fermion-fermion interaction is frequency independent this breaking of conservation laws does, however, not manifest in the lowest-order approximation with $m_{\rm c}=1$. In this, all interaction effects are encoded in renormalized single-particle parameters which become interaction, temperature, and bias voltage dependent during the RG flow \cite{Metzner2012,Karrasch2010}. This results in an effective single-particle picture. The self-energy is frequency independent and basic conservation laws hold to all orders \cite{Jakobs2010a,Jakobs2010}. For higher-order truncation schemes of Keldysh functional RG \cite{Jakobs2010a,Laakso2014,Kloeckner2020,Weidinger2021} the breaking of conservation laws might, however, have unwanted consequences even for small interactions, despite spoiling the results only to order $m_{\rm c}+1$ in the interaction. In the worst case, the results become useless. This depends on the details of the model studied as well as the parameter regime of interest and must be investigated on a case by case basis. We emphasize that formally the error is controlled to order ${m_{\rm c}}$ in the interaction and can thus be made small by reducing the strength of the two-particle interaction. However, at the same time, the interaction effect one is interested in might become unrecognizable. For recent methodological progress towards conserving approximations, see Refs.~\cite{Rentrop2016,Kugler2018,Kugler2019} and references therein.

When it comes to models with a frequency dependent (effective) two-particle interaction, such as the SAHM after integrating out the phonons, the breaking of conservation laws to quadratic order in the interaction might already be relevant in the lowest-order truncation. The inelastic processes (emission and absorption of phonons) lead to a frequency dependent self-energy and the effective single-particle picture does not apply any longer. Therefore, the SAHM, possibly complemented by a nearest-neighbor Coulomb repulsion of IRLM type, i.e., the CM, constitutes a model in which the role of the breaking of conservation laws can be investigated already in the lowest-order truncation. That is on a comparatively simple technical level and, thus, in a rather transparent way. We aim at exactly such an analysis. Although the lowest-order truncated functional RG scheme was used to investigate the SAHM in and out of equilibrium \cite{Khedri2017,Khedri2017b,Khedri2018}, no such critical evaluation of its limitations was given so far. The effects of the breaking of conservation laws for other models in higher-order truncations were discussed in Refs.~\onlinecite{Kloeckner2020,Weidinger2021}.    

Already at this stage, we would like to emphasize that on general grounds it is certainly desirable to devise approximations which obey fundamental conservation laws to all orders (conserving approximations). However, such approximations should not be employed at any price. If an approximation applied to, e.g., the IRLM does preserve conservation laws to all orders but does not capture the above mentioned power-laws with interaction dependent exponents, it might not be the method of choice. 

Take as a general example the mean-field approximation (self-consistent Hartree-Fock) which is conserving. In low-dimensional correlated systems (e.g. quantum dots and one-dimensional chains), however, it is prone to artificial spontaneous symmetry breakings. Applied to the Kondo problem it, e.g., leads to spurious breaking of the spin symmetry. For the problem at hand, it does not lead to a resummation of the series of leading logarithms. To reliably investigate many interesting issues of the correlation physics of low-dimensional systems mean-field theory should thus not be used. 

For a given model and set of observables, it will be difficult to a priori estimate the consequences of the breaking of conservation laws to order $m_{\rm c}+1$ in the two-particle interaction resulting out of the standard truncation of functional RG. This shows the importance of studies of the present type, in which we discuss how internal consistency checks can be used to evaluate the reliability of the results obtained by truncated functional RG. Our investigation might serve as a blueprint for the study of more complex models and/or the application of higher order truncations.  

The rest of this paper is structured as follows. In Sect.~\ref{sec:Modelandobservables} we introduce our model, provide basic steps for the solution, and give expressions for the transport properties of interest to us. We next present the truncated RG flow equations in Sect.~\ref{sec:funRG}. Details on the numerical solutions are given in the Appendix. In Sect.~\ref{sec:results} we discuss our results for $T_{\rm K}$, $I^{\rm N}(V)$, and the transport coefficients of the CM. The issue of current conservation is discussed in Sect.~\ref{sec:curcon}. A summary is presented in Sect.~\ref{sec:summary}.  

\section{Model and observables}
\label{sec:Modelandobservables}

\subsection{The Hamiltonian}

\begin{figure}[t]
	\center
	\tikzset{every picture/.style={line width=0.75pt}} 

\begin{tikzpicture}[x=0.75pt,y=0.75pt,yscale=-.9,xscale=.9]

\draw  [color={rgb, 255:red, 191; green, 208; blue, 209 }  ,draw opacity=1 ][fill={rgb, 255:red, 230; green, 236; blue, 236 }  ,fill opacity=1 ] (234.35,160.31) .. controls (234.35,153.09) and (240.32,147.23) .. (247.68,147.23) .. controls (255.04,147.23) and (261,153.09) .. (261,160.31) .. controls (261,167.54) and (255.04,173.4) .. (247.68,173.4) .. controls (240.32,173.4) and (234.35,167.54) .. (234.35,160.31) -- cycle ;
\draw  [color={rgb, 255:red, 191; green, 208; blue, 209 }  ,draw opacity=1 ][fill={rgb, 255:red, 230; green, 236; blue, 236 }  ,fill opacity=1 ] (151.36,128.29) -- (183.72,128.29) .. controls (201.6,128.29) and (216.08,142.67) .. (216.08,160.4) .. controls (216.08,178.14) and (201.6,192.51) .. (183.72,192.51) -- (151.36,192.51) -- cycle ;
\draw  [color={rgb, 255:red, 191; green, 208; blue, 209 }  ,draw opacity=1 ][fill={rgb, 255:red, 230; green, 236; blue, 236 }  ,fill opacity=1 ] (510.42,192.59) -- (478.06,192.55) .. controls (460.19,192.53) and (445.72,178.13) .. (445.74,160.4) .. controls (445.76,142.67) and (460.27,128.31) .. (478.14,128.33) -- (510.5,128.37) -- cycle ;
\draw  [color={rgb, 255:red, 100; green, 101; blue, 103 }  ,draw opacity=1 ][line width=1.5]  (344.29,100.47) .. controls (339.48,100.79) and (334.9,101.09) .. (334.42,102.75) .. controls (333.93,104.4) and (337.68,106.95) .. (341.61,109.61) .. controls (345.55,112.28) and (349.29,114.82) .. (348.81,116.48) .. controls (348.32,118.13) and (343.74,118.44) .. (338.93,118.76) .. controls (334.13,119.07) and (329.55,119.38) .. (329.06,121.04) .. controls (328.58,122.69) and (332.32,125.23) .. (336.26,127.9) .. controls (340.19,130.56) and (343.94,133.11) .. (343.45,134.76) .. controls (343.44,134.81) and (343.42,134.85) .. (343.4,134.9) ;
\draw [color={rgb, 255:red, 100; green, 101; blue, 103 }  ,draw opacity=1 ][fill={rgb, 255:red, 100; green, 101; blue, 103 }  ,fill opacity=1 ][line width=1.5]    (343.44,134.66) -- (334.66,144.21) ;
\draw [shift={(331.95,147.16)}, rotate = 312.58000000000004] [fill={rgb, 255:red, 100; green, 101; blue, 103 }  ,fill opacity=1 ][line width=0.08]  [draw opacity=0] (11.61,-5.58) -- (0,0) -- (11.61,5.58) -- cycle    ;
\draw [color={rgb, 255:red, 100; green, 101; blue, 103 }  ,draw opacity=1 ][fill={rgb, 255:red, 0; green, 0; blue, 0 }  ,fill opacity=0 ][line width=1.5]    (257.84,150.41) .. controls (279.24,139.66) and (300.12,139.16) .. (320.47,150.41) ;
\draw [color={rgb, 255:red, 100; green, 101; blue, 103 }  ,draw opacity=1 ][fill={rgb, 255:red, 0; green, 0; blue, 0 }  ,fill opacity=0 ][line width=1.5]    (341.35,150.41) .. controls (362.75,139.66) and (383.63,139.16) .. (403.98,150.41) ;
\draw [color={rgb, 255:red, 100; green, 101; blue, 103 }  ,draw opacity=1 ][fill={rgb, 255:red, 191; green, 208; blue, 209 }  ,fill opacity=1 ][line width=1.5]    (236.96,156.4) -- (250.34,156.25) -- (258.36,156.15) ;
\draw [color={rgb, 255:red, 100; green, 101; blue, 103 }  ,draw opacity=1 ][fill={rgb, 255:red, 0; green, 0; blue, 0 }  ,fill opacity=0 ][line width=1.5]    (216.08,150.41) .. controls (221.3,145.16) and (231.74,145.16) .. (236.96,150.41) ;
\draw [color={rgb, 255:red, 100; green, 101; blue, 103 }  ,draw opacity=1 ][fill={rgb, 255:red, 0; green, 0; blue, 0 }  ,fill opacity=0 ][line width=1.5]    (424.86,150.41) .. controls (430.08,145.16) and (440.52,145.16) .. (445.74,150.41) ;
\draw   (234.92,187.63) .. controls (234.9,192.3) and (237.22,194.64) .. (241.89,194.66) -- (321.57,195) .. controls (328.24,195.03) and (331.56,197.37) .. (331.54,202.04) .. controls (331.56,197.37) and (334.9,195.05) .. (341.57,195.08)(338.57,195.07) -- (421.25,195.41) .. controls (425.92,195.43) and (428.26,193.11) .. (428.27,188.44) ;
\draw   (317.21,169.43) .. controls (317.26,173.38) and (319.27,175.33) .. (323.22,175.27) -- (323.22,175.27) .. controls (328.86,175.19) and (331.71,177.13) .. (331.76,181.08) .. controls (331.71,177.13) and (334.5,175.11) .. (340.14,175.03)(337.6,175.07) -- (340.14,175.03) .. controls (344.09,174.98) and (346.03,172.98) .. (345.98,169.03) ;
\draw  [color={rgb, 255:red, 191; green, 208; blue, 209 }  ,draw opacity=1 ][fill={rgb, 255:red, 230; green, 236; blue, 236 }  ,fill opacity=1 ] (317.35,160.31) .. controls (317.35,153.09) and (323.32,147.23) .. (330.68,147.23) .. controls (338.04,147.23) and (344,153.09) .. (344,160.31) .. controls (344,167.54) and (338.04,173.4) .. (330.68,173.4) .. controls (323.32,173.4) and (317.35,167.54) .. (317.35,160.31) -- cycle ;
\draw [color={rgb, 255:red, 100; green, 101; blue, 103 }  ,draw opacity=1 ][fill={rgb, 255:red, 191; green, 208; blue, 209 }  ,fill opacity=1 ][line width=1.5]    (319.96,156.4) -- (333.34,156.25) -- (341.36,156.15) ;
\draw  [color={rgb, 255:red, 191; green, 208; blue, 209 }  ,draw opacity=1 ][fill={rgb, 255:red, 230; green, 236; blue, 236 }  ,fill opacity=1 ] (401.35,160.31) .. controls (401.35,153.09) and (407.32,147.23) .. (414.68,147.23) .. controls (422.04,147.23) and (428,153.09) .. (428,160.31) .. controls (428,167.54) and (422.04,173.4) .. (414.68,173.4) .. controls (407.32,173.4) and (401.35,167.54) .. (401.35,160.31) -- cycle ;
\draw [color={rgb, 255:red, 100; green, 101; blue, 103 }  ,draw opacity=1 ][fill={rgb, 255:red, 191; green, 208; blue, 209 }  ,fill opacity=1 ][line width=1.5]    (403.96,156.4) -- (417.34,156.25) -- (425.36,156.15) ;

\draw (274.11,123.31) node [anchor=north west][inner sep=0.75pt]    {$U,\ t$};
\draw (321.9,130.81) node [anchor=north west][inner sep=0.75pt]    {$\lambda $};
\draw (317.1,101.33) node [anchor=north west][inner sep=0.75pt]    {$\omega _{0}$};
\draw (161.51,152.41) node [anchor=north west][inner sep=0.75pt]    {$\mu _{L} ,T_{L}$};
\draw (239.96,158.81) node [anchor=north west][inner sep=0.75pt]    {$\epsilon _{1}$};
\draw (221.05,129.81) node [anchor=north west][inner sep=0.75pt]    {$\tau $};
\draw (429.82,130.31) node [anchor=north west][inner sep=0.75pt]    {$\tau $};
\draw (316,178.52) node [anchor=north west][inner sep=0.75pt]   [align=left] {\text{level}};
\draw (358.11,123.31) node [anchor=north west][inner sep=0.75pt]    {$U,\ t$};
\draw (455.51,152.41) node [anchor=north west][inner sep=0.75pt]    {$\mu _{R} ,T_{R}$};
\draw (322.96,158.81) node [anchor=north west][inner sep=0.75pt]    {$\epsilon _{2}$};
\draw (406.96,158.81) node [anchor=north west][inner sep=0.75pt]    {$\epsilon _{1}$};
\draw (320.43,199.52) node [anchor=north west][inner sep=0.75pt]   [align=left] {\text{dot}};

\end{tikzpicture}
	\caption{Schematics of the model describing a quantum dot with three energy levels of energy $\epsilon_j$ coupled to two reservoirs. The outer two dot levels are only added for computational reasons. Choosing $\tau \gg t$ and $\epsilon_1=0=\epsilon_3$ these two levels become part of the left and right reservoirs, respectively. The reservoirs have chemical potential $\mu_{L/R}$ and temperature $T_{L/R}$. Fermions occupying the three dot levels interact via a screened Coulomb interaction $U>0$. The central energy level is furthermore coupled by amplitude $\lambda$ to a phonon mode of frequency $\omega_0$.}
	\label{fig:Model}
\end{figure}
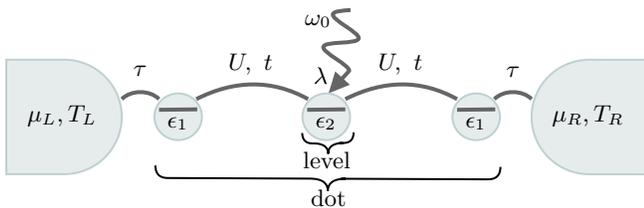

Our setup is sketched in Fig.~\ref{fig:Model}. We consider a central localized (impurity) level of energy $\epsilon_2$ symmetrically coupled to two noninteracting leads ($L$,$R$) at chemical potentials $\mu_{L/R}$ and temperatures $T_{L/R}$. The hopping on and of the impurity with matrix element $t$ is assumed to be much smaller than the scale of the lead bandwidth $D$. For computational reasons, we have to explicitly treat two levels at the boundaries of the reservoirs. They have energies $\epsilon_1=0=\epsilon_3$ and, by choosing the hopping $\tau$ to be of order $D$ and much larger than $t$, are effectively incorporated into the left and right reservoirs. We refer to the three levels combined as the dot. A voltage $V \geq 0$ is applied symmetrically across the dot which is included by setting $\mu_L=-\mu_R=V/2$.
The leads are taken in the thermodynamic limit. The fermions are assumed to be spinless and we are interested in the effect of two-particle interactions on the transport properties. We consider a screened, local Coulomb interaction $U \geq 0$ between fermions occupying the central impurity level and fermions located a the boundaries of the leads, i.e., between levels 1 and 2 as well as 2 and 3. In addition, we assume that with amplitude $\lambda \geq 0$ the occupation of the central level can lead to the emission or absorption of a phonon with frequency $\omega_0$. The phonon bath will be held at temperature $T_{\rm P}$. This model is, as mentioned above, a combination of the IRLM for $\lambda=0$ as well as the SAHM for $U=0$ and will be referred to as the CM. 

The model Hamiltonian consists of three parts: the dot, the leads, and the coupling between the two,
\begin{equation}
    H = H_\text{dot} + H_\text{res} + H_\text{coup}.
    \label{eq:hamfull}
\end{equation}
The dot part is split into a free one $H_\text{dot,0}$, containing only the three fermionic energy-levels and the tunneling between them (as well as $U$- and $\lambda$-dependent shifts of the zero of energy; see below), and the interacting part $H_\text{dot,int}$, describing the vibrational degree of freedom as well as the coupling to the central level and the Coulomb interaction. It takes the form
\begin{align}
	H_\text{dot,0} =& \left(\epsilon_1-\frac{U}{2}\right) (d_1^\dagger d_1 +d_3^\dagger d_3) + (\epsilon_2 - U +E_{\rm p}) d_2^\dagger d_2 \nonumber \\
	&+  t (d_1^\dagger d_2 + d_2^\dagger d_1 + d_2^\dagger d_3 + d_3^\dagger d_2)  \nonumber \\
	H_\text{dot,int} =& U (d_1^\dagger d_1 d_2^\dagger d_2 + d_2^\dagger d_2 d_3^\dagger d_3 ) \nonumber \\ 
	& + \lambda (b^\dagger + b) d_2^\dagger d_2 
	+\omega_0 b^\dagger b ,
	\label{eq:dotham}
\end{align}
with the polaronic shift $E_{\rm p} = \lambda^2/\omega_0$. Here $d_j$ and $d_j^\dagger$ are the fermionic annihilation and creation operators of the dot sites, and $b$ und $b^\dagger$ are the corresponding bosonic ones of the phonon mode. By assuming the leads to be particle-hole symmetric (PHS; see the next paragraph), the zero of energy is set such that the PHS point is given by $(\epsilon_1, \epsilon_2 , \epsilon_3)_\text{PHS}=(0, 0, 0)$.

The left ($s=L$) and right ($s=R$) reservoirs are both assumed to be non-interacting and given by
\begin{equation}
	H_\text{res} = \sum_{s=L,R} \sum_{k_s} \epsilon_{k_s} c_{k_s}^\dagger c_{k_s},
\end{equation}
where $c_{k_s}^\dagger$ and $c_{k_s}$ are the fermionic creation and annihilation operators for the lead states, respectively. PHS of the leads means that for each lead state with single-particle quantum number $k_s$ there exists a $k_s^\prime$ such that $\epsilon_{k_s}$ = $-\epsilon_{k_s^\prime}$, with the leads single-particle dispersion $\epsilon_{k_s}$.

Finally, the dot couples symmetrically to the leads with amplitude $\tau$
\begin{equation}
	H_\text{coup} = \frac{\tau}{\sqrt{N}} \sum_{k_L,k_R} \left[ d_1^\dagger c_{k_L} +c_{k_L}^\dagger d_1 +  d_3^\dagger c_{k_R} + c_{k_R}^\dagger d_3 \right].
\end{equation} 
We assumed reservoirs of size $N$ for the leads; in the following we consider $N \to \infty$. 

\subsection{The reservoir self-energy}

We are not interested in effects of the reservoir band structure on the transport properties and thus take the so-called wide-band limit; see, e.g., Ref.~\onlinecite{Karrasch2010}. In this the reservoirs are assumed to be structureless, i.e., the density of states becomes energy independent $\rho_\text{res}(\omega)=\rho_\text{res}$. Next, we can define a constant hybridization as $D=\pi\tau^2\rho_\text{res}$ which is the measure of the reservoir band width already considered above. 

When being interested in spectral properties of the central level or transport through the dot it is advantageous to integrate out the reservoir degrees of freedom via standard projection techniques \cite{Taylor1972}. This leads to a reservoir self-energy \cite{Jakobs2010a}. As we are aiming at finite bias properties we employ the formalism of Keldysh Green functions \cite{Rammer1986}. We here closely follow the steps outlined in detail in Refs.~\onlinecite{Karrasch2010} and \onlinecite{Khedri2018}. For the retarded, advanced and Keldysh components of the reservoir self-energy we obtain 
\begin{equation}
	\begin{aligned}
	\Sigma_\text{res}^\text{R/A}(\omega) 
	&= \pm iD  \begin{pmatrix}
	1&& \\ &0& \\ &&1
	\end{pmatrix},
	\\
	\Sigma_\text{res}^\text{K}(\omega)
	&=  - 2iD \begin{pmatrix}
	1-2f_L(\omega)&& \\ &0& \\ &&1-2f_R(\omega)
	\end{pmatrix},
	\end{aligned}
\end{equation}
where all matrix elements not written explicitly are zero. Here $f_{L/R}(\omega)$ denote the Fermi functions of the left and right reservoir respectively. We used the matrix notation in level-index space with $[\Sigma]_{jj'}= \Sigma_{jj'}$ for $j,j'\in\{1,2,3\}$.
The dots Green function is obtained by the Dyson equation as 
\begin{equation}
	G = \left[ G_0^{- 1} -\Sigma_\text{res} - \Sigma \right]^{- 1},
	\label{eq:dyson}
\end{equation}
with the noninteracting dot Green function $G_0$ and the self-energy $\Sigma$ due to the two-particle interactions.

\subsection{Integrating out the phonons}

As noted before, the bosonic degrees of freedom of the phonons can also be projected onto the dot, leading to an effective local, but purely retarded two-particle interaction for the fermions. Within the Keldysh formalism this interaction is of the form \cite{Laakso2014}
\begin{align}
	\bar{v}_{j_1^\prime,j_2^\prime|j_1,j_2}^{\nu_1^\prime,\nu_2^\prime|\nu_1,\nu_2}&  (\omega_1^\prime,\omega_2^\prime|\omega_1,\omega_2)  
	=  \delta(\omega_1^\prime+\omega_2^\prime-\omega_1-\omega_2) \nonumber \\
	& \times \frac{\lambda^2}{2\pi}
	\Big[ D^{\nu_1^\prime,\nu_2^\prime}(\omega_1^\prime-\omega_1) \delta_{\nu_1^\prime\nu_1}\delta_{\nu_2^\prime\nu_2}  \nonumber \\
	& \hspace{1cm} - D^{\nu_1^\prime,\nu_2^\prime}(\omega_1^\prime-\omega_2) \delta_{\nu_1^\prime\nu_2}\delta_{\nu_2^\prime\nu_1}\Big]  \nonumber \\
	& \times \sgn(\nu_1^\prime) \sgn(\nu_2^\prime) 
	\delta_{j_1^\prime=j_2^\prime=j_1=j_2=2}  
	.
	\label{eq:vphonon}
\end{align}
Here $\nu = \pm$ denotes the index of the Keldysh contour and 
$D(\omega)$ is the frequency-dependent phonon propagator. By assuming the steady state limit and utilizing particle-conservation, the retarded, advanced, and Keldysh component of the latter are given by
\begin{equation}
\begin{aligned}
	D^{\rm R}(\omega) &= D^{\rm A}(\omega)^*= \frac{1}{\omega-\omega_0+i\eta} - \frac{1}{\omega+\omega_0+i\eta}\\
	&= \frac{2\omega_0}{(\omega+i\eta)^2-\omega_0^2},
	\hspace{.5cm} \text{for } \eta \searrow 0,\\
	D^\text{K}(\omega) 
	& = -2i(1+2b(\omega_0)) \\&\times\left[ \frac{\eta}{(\omega+\omega_0)^2+\eta^2} + \frac{\eta}{(\omega-\omega_0)^2+\eta^2} \right] \\
	& \xrightarrow{\eta\rightarrow 0} -2\pi i\left[1+2b(\omega_0)\right] \left[ \delta(\omega+\omega_0)+\delta(\omega-\omega_0) \right],
\end{aligned}
\end{equation}
with the Bose-Einstein distribution $b(\omega)$ of the phonon bath with temperature $T_\text{p}$. We take $T_\text{p}=\frac{T_L+T_R}{2}$, which appears to be reasonable on physical grounds and does not further increase the number of parameters. The effective interaction is local in space, but not local in time. It is attractive and of order $\lambda^2$.

\subsection{Transport properties}
\label{sec:ParticleCurrent}

The particle current $I^{\rm N}$ through the dot is defined as the change of the number of particles in the leads. We distinguish between the left current flowing from lead $s=L$ into level $j=1$ and the right current from $s=R$ into $j=3$. Exploiting the steady-state as well as the wide-band limit, the current can be expressed in terms of the dot Green function as \cite{Meir1992} 
\begin{align}
    I^{\rm N}_s = &  - D \int_{-\infty}^\infty \dd\omega \text{Im}\left[ G_{jj}^\text{K}(\omega) - 2(1-2f_{s}(\omega)) G_{jj}^\text{R}(\omega) \right] \nonumber \\
    &\times(\delta_{s,L}\delta_{j,1}+\delta_{s,R}\delta_{j,3}).
\label{eq:ParticleCurrent}    
\end{align}
In equilibrium $V \to 0$ with $f_L(\omega)=f_R(\omega)=f(\omega)$, the dissipation-fluctuation theorem
\begin{align}
    \Im G^\text{K}(\omega) = 2\left[1-2f(\omega)\right]\Im G^\text{R}(\omega) 
    \label{eq:dissflu}    
\end{align}
holds and the integrand of Eq.~(\ref{eq:ParticleCurrent}) vanishes.

The Hamiltonian conserves the number of fermions which implies $I^{\rm N}_L=- I_R^{\rm N}$. The (symmetrized) current through the dot in the steady-state limit can thus also be written as
\begin{equation}
\label{eq:SymParticleCurrent}
    I^{\rm N} = \frac{I^{\rm N}_L - I^{\rm N}_R}{2}.
\end{equation}
However, if the Green functions are determined by an approximate scheme, e.g., a truncated functional RG approach, this conservation law can be violated. Therefore, it is useful to define the sum
\begin{equation}
	\Delta{I^{\rm N}} = |I^{\rm N}_L + I^{\rm N}_R|,
\end{equation}
which provides a measure of the violation.
As discussed before, the terms neglected by the first-order truncated functional RG are at least quadratic in the interaction strength. Thus $\Delta{I^{\rm N}}$ will be at least of second order in the amplitudes of the two-particle interaction $U$ and $\lambda^2$, i.e., of ${\mathcal O}\left( \left[ U,\lambda^2\right]^2 \right)$.

The particles also transport energy across the dot area. The steady-state energy current is defined as the change of energy in the reservoirs and given as (see, e.g., Ref.~\onlinecite{Kennes2013b})
\begin{align}
    I^{\rm E}_s =&
	- D \int_{-\infty}^\infty \!\!\!\! \dd\omega \, \omega \, \text{Im}\left\{ G_{jj}^\text{K}(\omega) - 2\left[1-2f_{s}(\omega)\right] G_{jj}^\text{R}(\omega) \right\} \nonumber \\
    & \times (\delta_{s,L}\delta_{j,1}+\delta_{s,R}\delta_{j,3}) .
\end{align}
From this and the particle current, the heat current can be computed as $I^{\rm Q}_s = I^{\rm E}_s-\mu_s I^{\rm N}_s$.

For finite $V$ and symmetrically applied temperature gradient $\Delta T=T_R-T_L$ we define generalized transport coefficients. 
This includes the conductance $G$, the Seebeck coefficient $S$, and the electron contribution to the thermal conductance $\kappa_e$ given by \cite{Onsager1931}
\begin{align}
    G &= \partial_V I^{\rm N} \big|_{\Delta T}, \label{eq:lin_con}\\
    S &= -\frac{\partial_{\Delta T}I^{\rm N}\big|_{V}}{\partial_V I^{\rm N} \big|_{\Delta T}}, \label{eq:S_def} \\
    \kappa_{\rm e} &= \partial_{\Delta T} I^{\rm Q}_R \big|_{V} 
    -\partial_{\Delta T} I^{\rm Q}_R \big|_{V} \; \frac{\partial_{\Delta T}I^{\rm N}\big|_{V}}{\partial_V I^{\rm N} \big|_{\Delta T}}. 
    \label{eq:ThermalConductance}
\end{align}

\section{Functional RG flow equations}
\label{sec:funRG}

To treat both the explicit fermion-fermion interaction as well as the retarded effective interaction resulting from integrating out the phonons we use the one-particle irreducible functional RG approach \cite{Metzner2012} on the Keldysh contour \cite{Jakobs2007} already introduced in Sec.~\ref{sec:intro}.  

The low-energy properties are regularized by introducing an infrared cutoff parameter $\Lambda$ into the noninteracting dot propagator which during the RG flow is sent from infinity down to zero. This way the RG idea of treating energy scales successively is introduced. For $\Lambda=0$ the cutoff-free problem of interest is recovered. 

Here the cutoff is realized by auxiliary leads tunnel-coupled with an amplitude $\Lambda$ to the three dot levels. At the end of the flow, at $\Lambda=0$, the auxiliary leads are decoupled \cite{Jakobs2010a}. In equilibrium this cutoff procedure preserves the dissipation-fluctuation theorem Eq.~(\ref{eq:dissflu}) \cite{Jakobs2010}. Even after picking this cutoff, we are free to select the auxiliary lead distribution functions. A proper selection requires physical guidance. The auxiliary reservoirs coupled to the first and third level are constructed to resemble the (physical) left and right lead. We thus assume the distribution function of these auxiliary leads to be $f_L(\omega)$ and $f_R(\omega)$, respectively. This way the total reservoir-level coupling to the dot levels 1 and 3 is given by the sum of the physical coupling $D$ and $\Lambda$. The total coupling flows from $\infty$ down to $D$ at the end of the RG procedure. The reservoir coupled to the central level is assumed to have the symmetrized distribution function $\frac{f_L(\omega) + f_R(\omega)}{2}$ \footnote{Other choices of auxiliary lead distribution functions than this physically intuitive one are conceivable. We have investigated different choices, but the physically motivated one taken in this work seems to be the natural choice for the problem at hand.}. Taking the derivative of the generating functional of one-particle irreducible vertex functions with respect to $\Lambda$ and expanding with respect to powers of the fields one obtains an infinite hierarchy of coupled differential (flow) equations for the vertex functions.  

We note that without truncation the final result for the vertex functions would be exact and, in particular, independent of the choice of the cutoff procedure. However, for practical computations truncations are required. We follow the procedure described in the Introduction to obtain a finite (closed) set of equations. After truncation at order $m_{\rm c}$, the vertex functions might acquire a dependence on the cutoff procedure to order $m_{\rm c}+1$ in the interaction \cite{Kloeckner2020}. In our concrete realizations this, e.g., implies that the self-energy (single-particle vertex) obtained for different choices of the auxiliary reservoir distribution functions show differences to order $m_{\rm c}+1$. It depends on the details of the problem of interest and the observables computed if this spoils the results to an extent, such that they become useless. In analogy to the role of broken current conservation, investigating this cutoff dependence would also require a case by case study. Experience shows that selecting the cutoff procedure in a physically plausible way (for the present setup as, e.g., described above) can prevent severe artifacts. Any further discussion of the cutoff dependence of the truncated functional RG results is beyond the scope of the present paper.   

By setting the two-particle vertex containing both interactions to its bare value, we truncate the hierarchy after first order, $m_{\rm c}=1$. This leads to a set of differential equations for the self-energy components (in level-index and Keldysh-index space). Details on the derivation can be found in Refs.~\onlinecite{Karrasch2010,Khedri2018}. Using general symmetries of the self-energies as well as those specific to the model under consideration, the flow equations can be brought into the form
\begin{equation}
\label{eq:FlowEquation}
\begin{aligned}
	\dot{\Sigma}^{\text{R},\Lambda}_{11} = 	\dot{\Sigma}^{\text{R},\Lambda}_{33} = & - \frac{U}{4\pi}\int \dd{\omega_1} \Im S^{\text{K},\Lambda}_{22}(\omega_1), \\
	\dot{\Sigma}^{\text{R},\Lambda}_{12} = \dot{\Sigma}^{\text{R},\Lambda}_{21} 
	= & \frac{U}{4\pi}\int \dd{\omega_1} \Im S^{\text{K},\Lambda}_{12}(\omega_1), \\
	\dot{\Sigma}^{\text{R},\Lambda}_{23} = \dot{\Sigma}^{\text{R},\Lambda}_{32} 
	= & \frac{U}{4\pi}\int \dd{\omega_1} \Im S^{\text{K},\Lambda}_{23}(\omega_1), \\
	\Re\dot{\Sigma}^{\text{R},\Lambda}_{22}(\omega) = &
	- \frac{U}{4\pi}\int \dd{\omega_1} [\Im S^{\text{K},\Lambda}_{11}(\omega_1)
	+\Im S^{\text{K},\Lambda}_{33}(\omega_1)] \\
	& + \frac{\lambda^2}{2\pi\omega_0} \int \dd{\omega_1} \Im S^{\text{K},\Lambda}_{22}(\omega_1) \\
	& -\frac{\lambda^2}{2} \sum_{s=\pm} 
	\Bigg\{ (1+2b(\omega_0)) \Re S_{22}^{\text{R},\Lambda}(\omega-s\omega_0) \\
	&- \frac{1}{2\pi} \int \!\!\!\!\!\! {\mathcal P} \dd{\omega_1} \left[ \frac{s\Im S_{22}^{\text{K},\Lambda}(\omega_1)}{\omega_1-\omega+s\omega_0} \right] \Bigg\}, \\
	\Im\dot{\Sigma}^{\text{R},\Lambda}_{22}(\omega) = & - \frac{\lambda^2}{2} \sum_{s=\pm} 
	\Big[ (1+2b(\omega_0))\Im S_{22}^{\text{R},\Lambda}(\omega-s\omega_0) \\
	& + \frac{s}{2} \Im S_{22}^{\text{K},\Lambda}(\omega-s\omega_0) \Big],\\
	\Im\dot{\Sigma}^{\text{K},\Lambda}_{22}(\omega) = & - \frac{\lambda^2}{2} \sum_{s= \pm}
	\Big[ 2s\Im S_{22}^{\text{R},\Lambda}(\omega-s\omega_0) \\
	& - (1+2b(\omega_0))\Im S_{22}^{\text{K},\Lambda}(\omega-s\omega_0) \Big],
\end{aligned}
\end{equation}
with $\mathcal P$ indicating the Cauchy principal value and the (single-scale) propagator is given by
\begin{equation}
\begin{aligned}
	S^\text{R}(\omega) 
	=& \left[ S^\text{A}(\omega) \right]^\dagger = i\left[ G^\text{R}(\omega) \right]^2,\\
	S^\text{K} (\omega) 
	=& iG^{\text{R},\Lambda}(\omega) G^{\text{K},\Lambda}(\omega)
	-iG^{\text{K},\Lambda}(\omega) G^{\text{A},\Lambda}(\omega) \\
	& + 2i G^{\text{R},\Lambda}(\omega) \left[\mathds{1}-2F_\text{aux}(\omega)\right]G^{\text{A},\Lambda}(\omega).
\end{aligned}
\end{equation}
The matrix $F_\text{aux}(\omega)$ contains the distribution functions of the auxiliary reservoirs and is defined as
\begin{align}
    F_\text{aux} =
    \begin{pmatrix}
        f_L(\omega) & & \\
         & \frac{f_L(\omega)+f_R(\omega)}{2} & \\
         & & f_R(\omega) 
    \end{pmatrix} .
\end{align}
The initial values of the components of the self-energy are 
\begin{equation}
\begin{aligned}
    \lim_{\Lambda\rightarrow\infty}\Sigma_{11}^{\text{R},\Lambda}= \lim_{\Lambda\rightarrow\infty}\Sigma_{33}^{\text{R},\Lambda}
    & =0, \\
    \lim_{\Lambda\rightarrow\infty}\Sigma_{12}^{\text{R},\Lambda}
    =\lim_{\Lambda\rightarrow\infty}\Sigma_{21}^{\text{R},\Lambda}&=t, \\
    \lim_{\Lambda\rightarrow\infty}\Sigma_{23}^{\text{R},\Lambda}
    =\lim_{\Lambda\rightarrow\infty}\Sigma_{32}^{\text{R},\Lambda}&=t, \\
    \lim_{\Lambda\rightarrow\infty}\Re\Sigma_{22}^{\text{R},\Lambda}&=\epsilon_2, \\ 
    \lim_{\Lambda\rightarrow\infty}\Im\Sigma_{22}^{\text{R},\Lambda}
    =\lim_{\Lambda\rightarrow\infty}\Im\Sigma_{22}^{\text{K},\Lambda}&=0.
\end{aligned}
\end{equation}
All results obtained from integrating this set of coupled differential equations from $\Lambda=\infty$ down to $\Lambda=0$ are exact (at least) up to $\mathcal{O}(U)$ and $\mathcal{O}(\lambda^2)$. For general parameters, the system of equations is too complex for an analytical solution. However, in the IRLM limit $\lambda=0$, analytical insights can be gained and were discussed in Refs.~\onlinecite{Karrasch2010,Karrasch2010b,Kennes2013}. 

We emphasize that the explicit frequency dependence of the retarded fermion-fermion interaction resulting from integrating out the phonons leads to a frequency-dependent self-energy even in the lowest-order truncation. The effective single-particle picture, which is very useful in the $\lambda=0$ IRLM limit is no longer applicable. Conservation laws, such as, e.g., current conservation are warranted to leading order only. Due to this frequency dependence, even the numerical solution of the flow equations becomes a challenge. Details on this can be found in the Appendix.

\section{Results}
\label{sec:results}

\subsection{The emergent low-energy energy scale}

The energy scale characterizing the coupling between the central dot level $j=2$ and the rest of the system for $U=0=\lambda$ is 
\begin{equation}
	\Gamma= \frac{2 t^2}{D} .
	\label{eq:Gammadef}    
\end{equation}
We are exclusively interested in the scaling limit with $\Gamma/D \ll 1$.
Including the interactions, the energy scale $\Gamma$ is renormalized and denoted as $T_{\rm K}$. This emergent low-energy scale can be defined via the zero temperature slope of the occupancy $n_2$ of the central level with $\epsilon_2$ in equilibrium \cite{Karrasch2010}
\begin{equation}
	T_{\rm K} = \frac{1}{\pi \chi}
	\text{, \hspace{.5cm}with } \chi = -\frac{dn_2}{d\epsilon_2} \bigg|_{\epsilon_2=V=T_L=T_R=0}. 
	\label{eq:T_K}
\end{equation}
Note the additional factor of 2 in the definition of $T_{\rm K}$ in, e.g., Ref.~\onlinecite{Karrasch2010}. Computing the derivative numerically based on Keldysh functional RG data for $n_2$ did turn out not to be stable. Here we therefore use the alternative definition of $T_{\rm K}$ as the half width at half maximum of the single-particle spectral function of the level $j=2$ (which we obtain easily in Keldysh functional RG, see, e.g., Ref.~\onlinecite{Khedri2018}, as compared to Matsubara functional RG \cite{Metzner2012}). 

An analytical expression for $T_{\rm K}$ of the SAHM obtained within lowest-order truncated Matsubara functional RG was presented in Ref.~\onlinecite{Khedri2017}. The expression resulting in the antiadiabatic limit of interest to us is given in Eq.~(\ref{eq:SAHMscale}). The expression for $T_{\rm K}$ of the IRLM derived by this approach is given in Eq.~(\ref{eq:IRLMscale}) with $\alpha_\Gamma= 2U/(\pi D)$ \cite{Karrasch2010b}. 

It can be shown using Matsubara functional RG and the same approximations employed in Ref.~\onlinecite{Khedri2017} to obtain $T^{\rm SAHM}_{\rm K}$, that, up to corrections of order ${\mathcal O} \left( \left[ U, \lambda^2 \right]^2 \right)$, the $T^{\rm CM}_{\rm K}$ of the combined model is just the product of that of the IRLM and the SAHM 
\begin{equation}
	\frac{T_{\rm K}^\text{CM}}{\Gamma} 
	= \frac{T_{\rm K}^\text{IRLM}}{\Gamma}\; \frac{T_{\rm K}^\text{SAHM}}{\Gamma} .
\end{equation}
Note that that while a $U>0$ increases the level-lead coupling, i.e., $T_{\rm K}^\text{IRLM} > \Gamma$, $\lambda^2>0$ leads to a reduction $T_{\rm K}^\text{IRLM} < \Gamma$. The two interactions compete in the CM and $\frac{T_{\rm K}^\text{CM}}{\Gamma} $ can become larger or smaller than 1 depending on which interaction prevails. 

Next, we will investigate if the results for the finite bias current and the transport coefficients of the CM can also be understood as superpositions of the characteristics found for the IRLM and the SAHM discussed in the introduction. 

\subsection{The fermionic particle current}

\begin{figure}[t]
	\centering
	\includegraphics[width=1\linewidth]{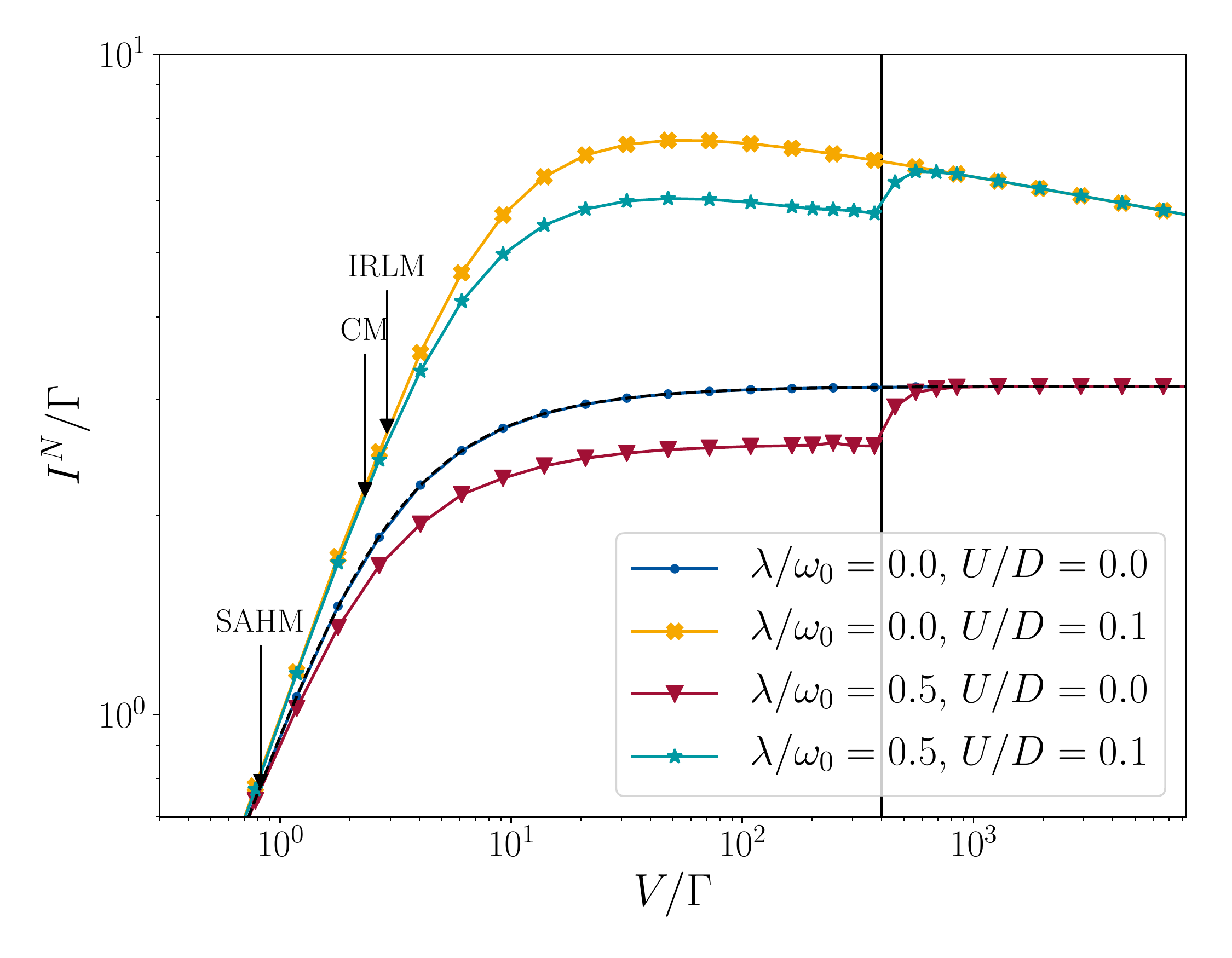}
	\caption{The particle current $I^{\rm N}$ as a function of the voltage $V$ for different interaction strength $\lambda$ and $U$ on a log-log scale. We consider the antiadiabatic limit $\omega_0/\Gamma=200$, particle-hole symmetry $\epsilon_2=0$, and vanishing temperature $T_L=T_R=T_\text{p}=0$. With $\Gamma/D=2\cdot 10^{-8}$ the scaling limit is realized. The black vertical line indicates $V=2\omega_0$ and the black, dotted curve, mostly hidden by the blue one, shows the analytically calculated noninteracting current. The vertical arrows indicate the three Kondo temperatures $T_{\rm K}^\text{CM}$, $T_{\rm K}^\text{IRLM}$, and $T_{\rm K}^\text{SAHM}$. At particle-hole symmetry the current is conserved and $\Delta I^{\rm N}=0$.}
	\label{fig:IN_step}
\end{figure}

\begin{figure}[t]
	\centering
	\includegraphics[width=1\linewidth]{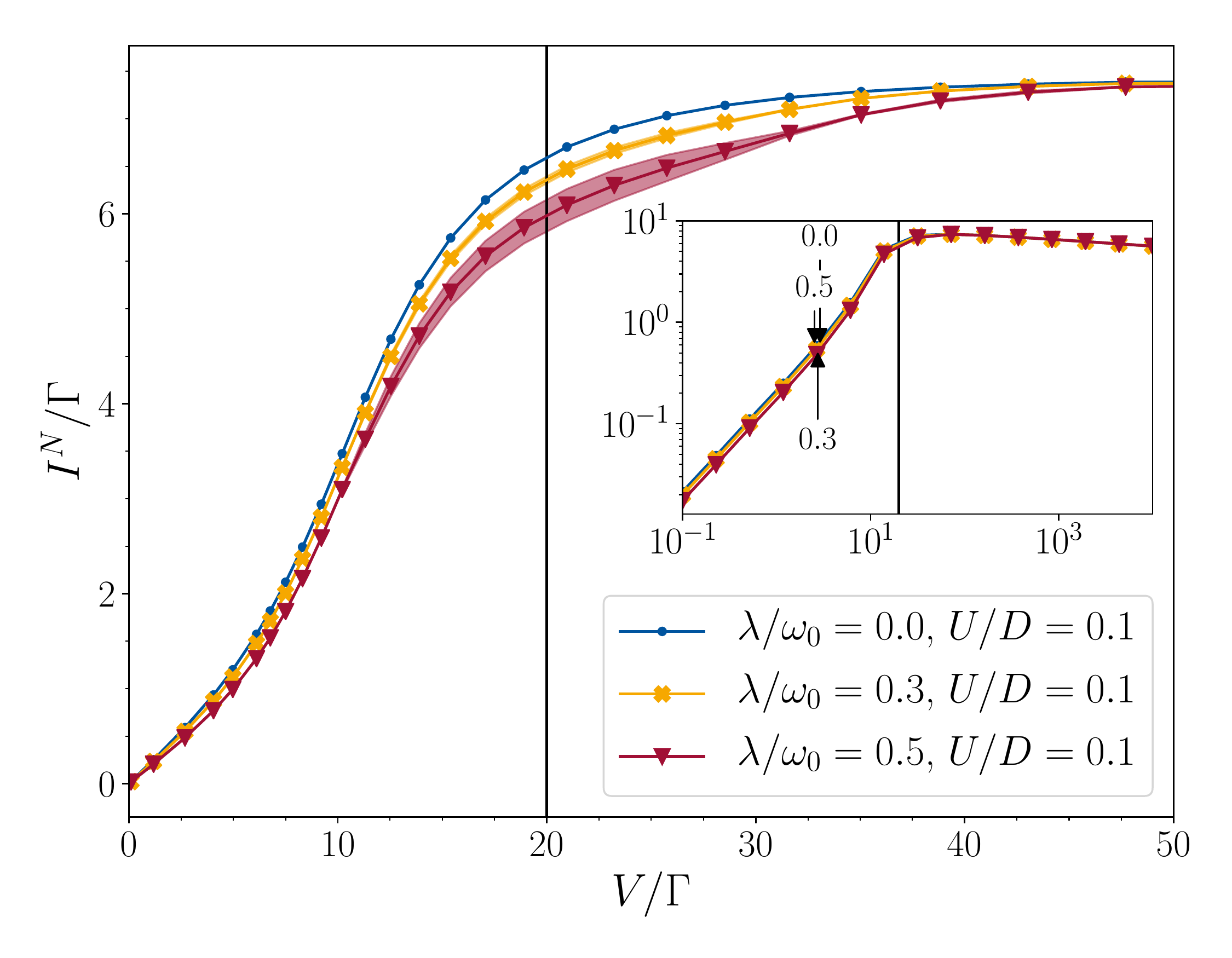}
	\caption{Main panel: The same as in Fig.~\ref{fig:IN_step}, but away from particle-hole symmetry, $\epsilon_2/\Gamma=5$, for $\omega_0/\Gamma=10$, and fixed $U$. A linear $x$- and $y$-axis scale is chosen. The measure for the broken current conservation $\Delta I^{\rm N}$ is indicated as the filled area.
	The black vertical line indicates $V=2\omega_0$. The inset shows a larger voltage-range to outline the position of the phononic step compared to the one of the power law decay. Note the logarithmic axes of the inset. The vertical arrows indicate $T_K$ for the different values of $\lambda/\omega_0$. For a discussion on this, see Sect. ~\ref{sec:curcon}.}
	\label{fig:IN_step_2}
\end{figure}

To study the interplay of the Coulomb and the fermion-phonon interaction in observables, we first computed the fermionic particle current through the interacting dot-region. In Fig.~\ref{fig:IN_step}, we show the $V$-dependence of $I^{\rm N}$ for a wide range of voltages at the particle-hole symmetric point $\epsilon_2=0$ and deep in the antiadiabatic regime with $\omega_0/\Gamma=200$. Note the logarithmic scales of the $x$- and $y$-axis. Different combinations of Coulomb- and fermion-phonon interactions were considered. The black, dotted line, which is mostly hidden by the blue one, shows the analytically determined interaction-free result
\begin{equation}
	\frac{I^N}{\Gamma} = 
	\tan^{-1}\left[ \frac{\epsilon_2+V/2}{\Gamma} \right] 
	- \tan^{-1}\left[ \frac{\epsilon_2-V/2}{\Gamma} \right]
\end{equation}
for reference. The main feature of these current-voltage characteristics for $\lambda >0$ is a phonon step at voltage $V=2 \omega_0$. At this voltage phonon satellites in the single-particle spectral function enter the transport window \cite{Khedri2018}. The step is robust against the inclusion of the Coulomb interaction. For $U>0$ we find the power-law suppression Eq.~(\ref{eq:currIRLM}) for $T_{\rm K} \ll V \ll D$; straight line on the log-log scale of Fig.~\ref{fig:IN_step}. The three Kondo temperatures $T_{\rm K}^\text{CM}$, $T_{\rm K}^\text{IRLM}$, and $T_{\rm K}^\text{SAHM}$ are indicated by vertical arrows. This suppression is also found if in addition $\lambda>0$. As $\omega_0 > T_{\rm K}^\text{CM}$ the suppression already sets in before the phonon step leads to a parallel shift of the current. We note that only after including the phonon satellite into the transport window the current of the CM reaches the same value as that of the corresponding IRLM ($\lambda=0$). For the parameters of Fig.~\ref{fig:IN_step} the characteristics of the Coulomb interaction and the fermion-phonon interaction occur on different scales and are thus superimposed in a straightforward way.  

To further investigate this, in Fig.~\ref{fig:IN_step_2} we show (in the main panel on linear $x$- and $y$-axis scales) the current-voltage characteristics away from particle-hole symmetry and for parameters in which $T_{\rm K}^{\rm CM}$ and $\omega_0$ are less well separated; $\omega_0/\Gamma=10$. For increasing $\lambda$ the current at small voltages is suppressed. This is the Frank-Condon blockade. For the present parameters the phonon step is more shoulder-like and for $\lambda=0.3$ only barely visible. The inset shows the same data on a log-log-scale. The onset of the IRLM-like power-law suppression is rather close to the position of the phonon shoulder at $V \approx 2 \omega_0$ but the interplay of both interaction effects does still not lead to any additional features. In the main panel, the measure for the broken current conservation $\Delta I^{\rm N}$ is indicated as the filled area. For a discussion on this, see Sect. ~\ref{sec:curcon}. 

\subsection{Transport coefficients}

\subsubsection{Linear transport}

\begin{figure}[t]
	\centering
	\includegraphics[width=1\linewidth]{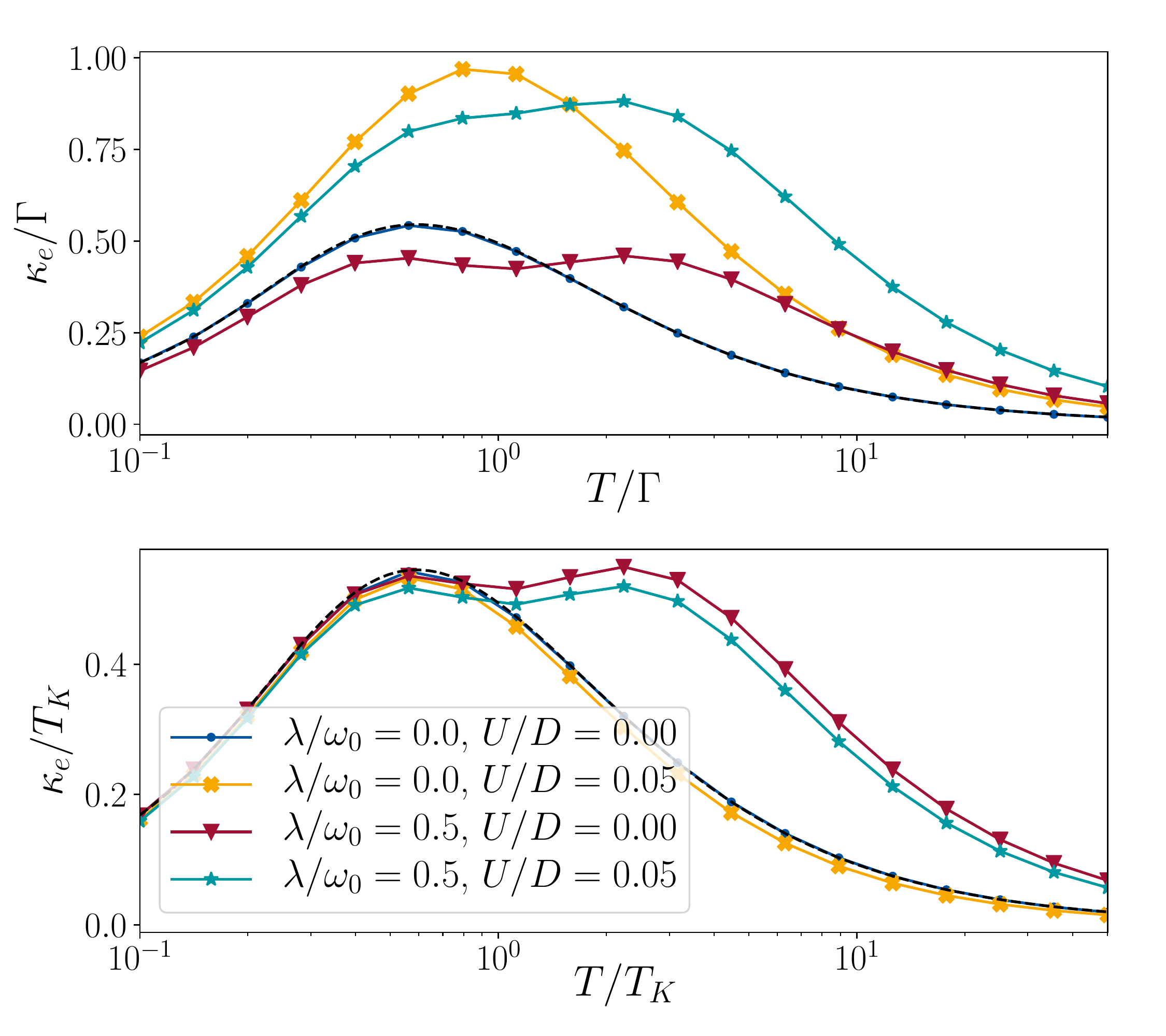}
	\caption{The temperature dependence of the electron contribution to the linear thermal conductance $\kappa_e$ in equilibrium $V/\Gamma=0$. The upper panel shows $\kappa_e$ as well as $T$ on the scale of the bare hybridisation $\Gamma$. $\epsilon_2/\Gamma=-1$ and $\omega_0/\Gamma=5$ are also given in units of $\Gamma$. The dotted line shows the analytically calculated free case. In the lower panel $T$ and $\kappa_{\rm e}$ are given in units of $T_{\rm K}^{\rm CM}$. The phonon frequency and the level position were changed such that $\epsilon_2/T_{\rm K}^{\rm CM}=-1$ and $\omega_0/T_{\rm K}^{\rm CM}=5$. The curves for different $U$ collapse.}
	\label{fig:kappa_renorm_V0}
\end{figure}

\begin{figure}[t]
	\centering
	\includegraphics[width=1\linewidth]{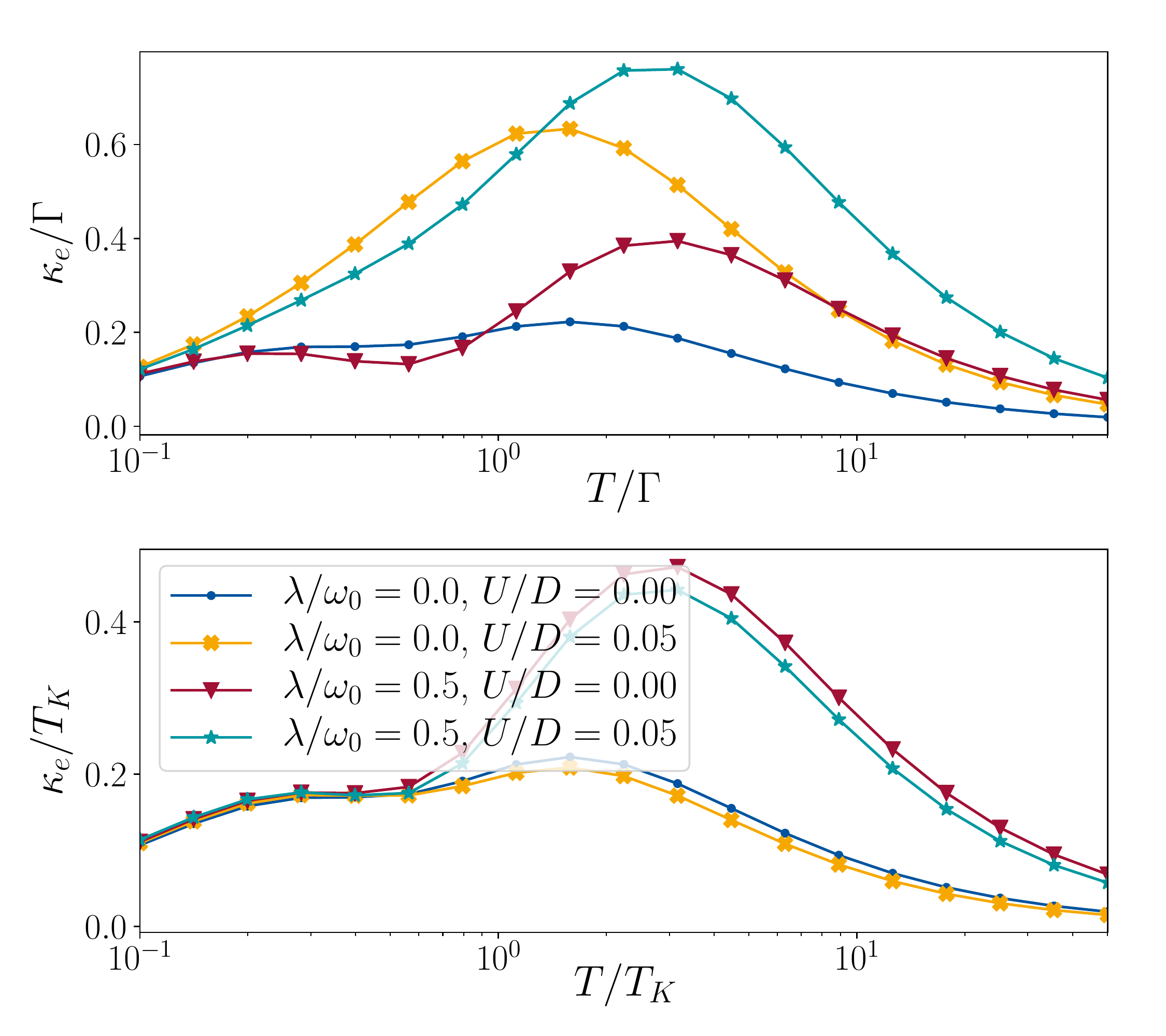}
	\caption{The temperature dependence of the electron contribution to the thermal conductance $\kappa_e$ away from equilibrium. The upper plot shows $\kappa_e$ as well as $T$ as a function of the bare hybridisation $\Gamma$. $\epsilon_2/\Gamma=-1$, $\omega_0/\Gamma=5$ and $V/\Gamma=2$ are also scaled with $\Gamma$. The lower plot renormalises the small energy scales with $T_K = T_K(\omega_0/\Gamma=5)$ instead, so that $\epsilon_2/T_K=-1$, $\omega_0/T_K=5$ and $V/T_K=2$.}
	\label{fig:kappa_renorm_V2}
\end{figure}

We will next focus on the linear transport coefficients, with $V \to 0$ und $T_L \to T_R=T$. It turns out that the effects of the two interactions on these observables in the combined model can be understood as the superposition of the renormalization of the level-lead tunneling and the $U=0$ phonon features of the SAHM emerging due to inelastic scattering processes. As the electron contribution to the thermal conductance $\kappa_{\rm e}$ Eq.~(\ref{eq:ThermalConductance}) shows the most interesting behavior in the SAHM limit \cite{Khedri2017b}, we here focus on this. In contrast to the other transport coefficients, it furthermore contains information on the particle and heat current as well as on the changes with voltage and temperature difference of the leads. We already now note that this superposition also holds for the other coefficients Eqs.~(\ref{eq:lin_con}) and (\ref{eq:S_def}), i.e., the linear conductance and the Seebeck coefficient.  

Results for $\kappa_{\rm e}$ at different interaction strength as a function of the temperature $T$ are shown in Fig.~\ref{fig:kappa_renorm_V0}. The other parameters are $\epsilon_2/\Gamma=-1$ and $\omega_0/\Gamma=5$. For reference, we show the noninteracting result, which can analytically be expressed in terms of polygamma functions \cite{Murphy2008}, as a black dotted line. It is mostly hidden by the blue line. For $\lambda=0$, we observe a maximum whose position we found to scale with $\epsilon_2$ and which thus, originates from resonant tunneling. A second peak emerges when introducing a finite $\lambda$. It scales with $\omega_0$ and is therefore linked to phononic resonances. In the upper panel, $T$ and $\kappa_e$ are both depicted in units of $\Gamma$. We next computed $T_{\rm K}^{\rm CM}$ for the respective interaction strengths. We then used this as our unit of energy instead of $\Gamma$, i.e., we recomputed $\kappa_{\rm e}$ for $\epsilon_2/T_{\rm K}^{\rm CM}=-1$ and $\omega_0/T_{\rm K}^{\rm CM}=5$. If $T$ and $\kappa_{\rm e}$ are in addition rescaled by $T_{\rm K}^{\rm CM}$ this leads to the collapse of all curves for a fixed $\lambda$ but with different $U$, including $U=0$, onto a single curve as shown in the lower panel of Fig.~\ref{fig:kappa_renorm_V0}. In that sense $T_{\rm K}^{\rm CM}$ acts as a universal low-energy scale. The effect of the Coulomb interaction on $\kappa_{\rm e}$ of the combined model can, in particular, be fully incorporated into the renormalization of $T_{\rm K}^{\rm CM}$. The same type of collapse can be reached for the other transport coefficients. We note that the Kondo scale of the combined model depends on $\omega_0$. Completely following the above logic of finding the right parameters for a collapse of the curves for different $U$ would thus require an iterative process. However, the corrections would be of higher order in the interaction and can be neglected.  

\subsubsection{Nonlinear transport}

We now show that the collapse of the transport coefficients for different $U$ on a $\lambda$-dependent curve can be achieved also beyond the linear response regime \cite{Kennes2013b}. In Fig.~\ref{fig:kappa_renorm_V2} the same as in Fig.~\ref{fig:kappa_renorm_V0} is shown, but for $V/\Gamma=2$ (upper panel) and $V/T_{\rm K}^{\rm CM}=2$ (lower panel). Comparing to Fig.~\ref{fig:kappa_renorm_V0} shows that the finite bias voltage modifies the resonant tunneling peak but only has a weak effect on the phonon resonance. Regardless of this, the collapse works equally well as compared to the linear response regime (see the lower panel). Again the same holds for the other transport coefficients.
 
\section{Current conservation}
\label{sec:curcon}

Even though the Hamiltonian Eq.~(\ref{eq:hamfull}) is particle conserving, the truncation of the flow equations can lead to a violation of this conservation and thus to $\Delta I^{\rm N} > 0$. In Sect.~\ref{sec:results}, we have focused on parts of the parameter space in which this effect is negligible. In particular, at PHS, i.e. for $\epsilon_2=0$, $\Delta I^{\rm N} = 0$ holds regardless of the other parameters. But even away from PHS, we identified parameter regimes with interesting nonequilibrium many-body physics in which $\Delta I^{\rm N}$ can be neglected. This can, e.g., be seen from the filled area in the main panel of Fig.~\ref{fig:IN_step_2} which indicates $\Delta I^{\rm N}$. 

\begin{figure}[t]
	\centering
	\includegraphics[width=1\linewidth]{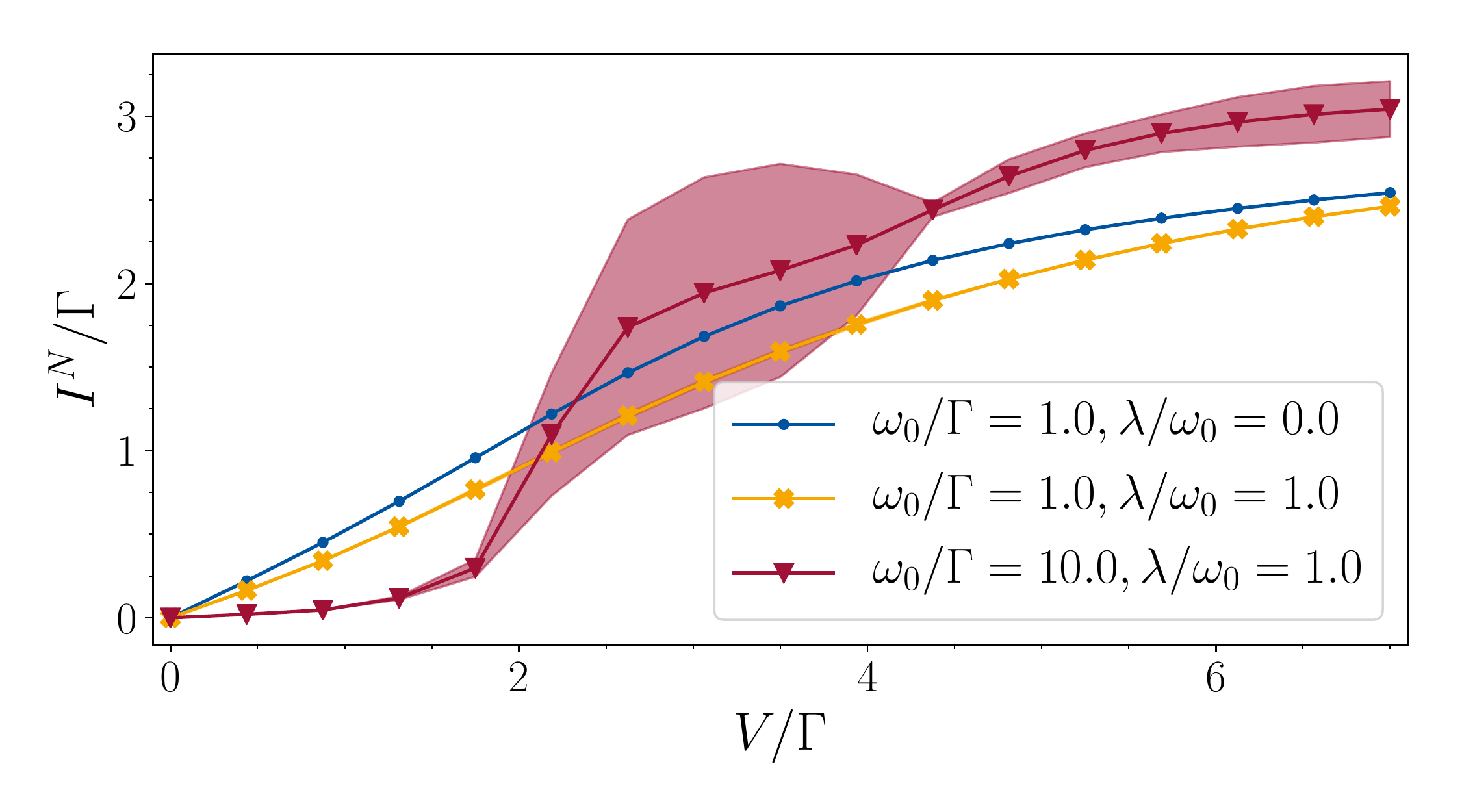}
	\caption{The particle current $I^{\rm N}$ as a function of the voltage $V$. The SAHM limit $U=0$ for different $\lambda$ and $\omega_0$ is shown. The filled area indicates the violation of current conservation $\Delta I^{\rm N}$. The temperature is $T_L=T_R=T_\text{p}=0.01\omega_0$ and the on-site energy $\epsilon_2/\omega_0=0.01$.}
	\label{fig:IN_largeDeltaIN}
\end{figure}

\begin{figure}[t]
	\centering
	\includegraphics[width=1\linewidth]{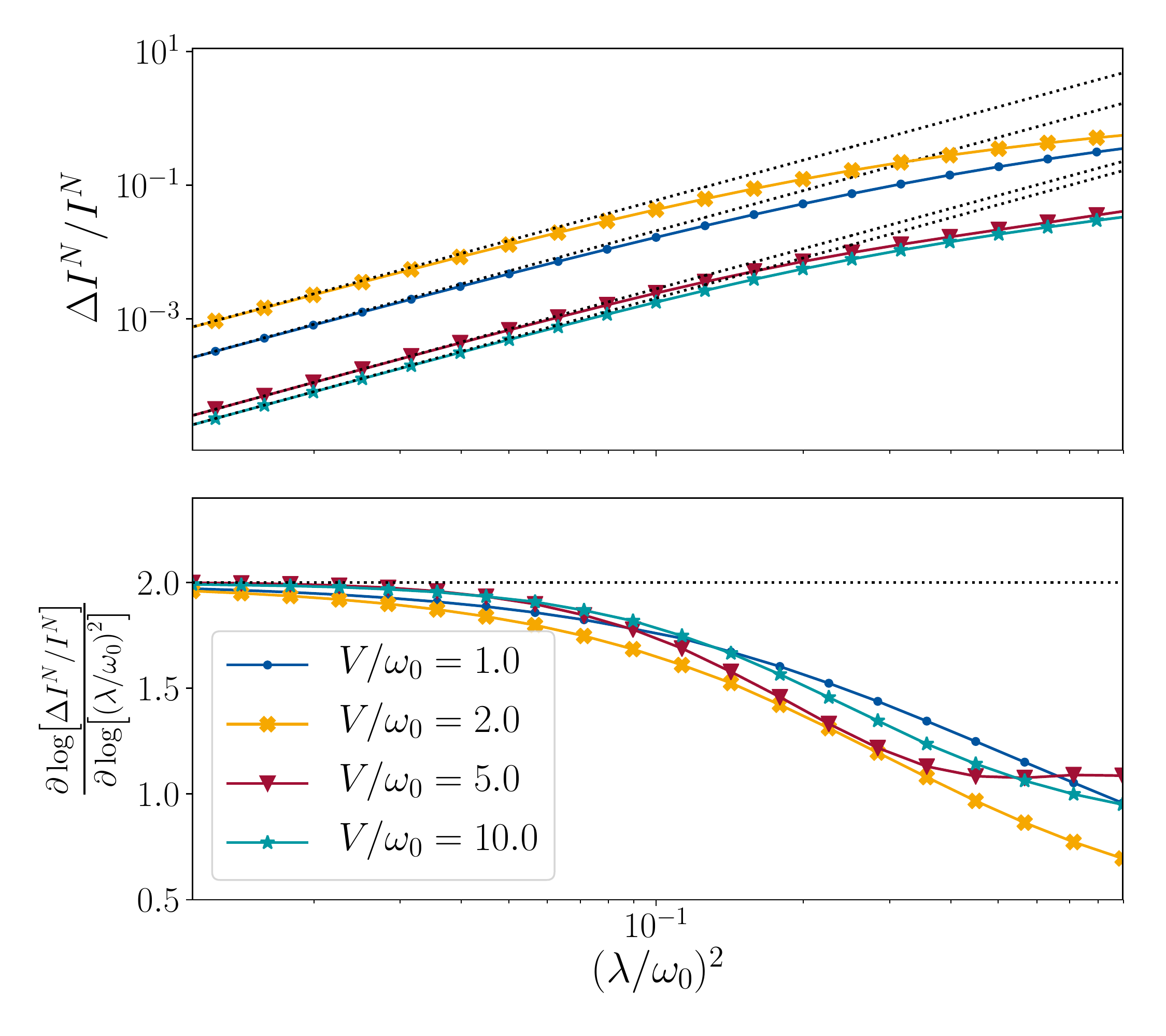}
	\caption{Relative violation of the current conservation $\Delta I^{\rm N}/I^{\rm N}$ as a function of the interaction strength $(\lambda/\omega_0)^2$ for $U=0$ in the anti-adiabatic limit $\omega_0/\Gamma=10$ and away from particle-hole-symmetry $\epsilon_2/\Gamma=5$. The temperature is $T_L=T_R=\Gamma$ is assumed and various bias-voltage $V$ are considered. The dotted lines indicate the expected power laws $\sim (\lambda/\omega_0)^4$ for comparison. Upper panel: bare data on a log-log scale. Lower panel: logarithmic derivative (apparent exponent).}
	\label{fig:DeltaIN_QuadLams_Vs}
\end{figure}

\begin{figure}[t]
	\centering
	\includegraphics[width=1.\linewidth]{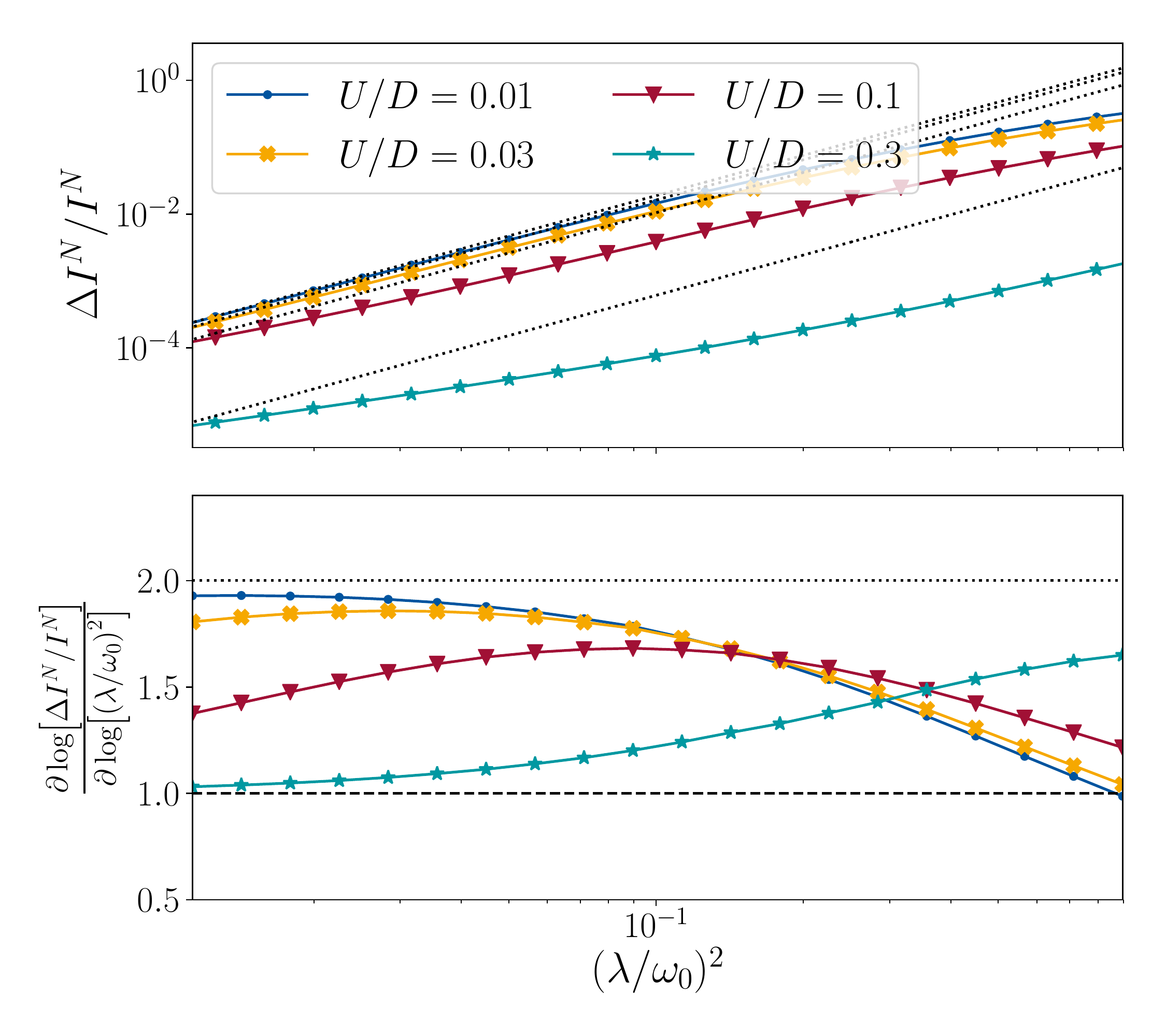}
	\caption{Relative violation of the current conservation $\Delta I^{\rm N}/I^{\rm N}$ as a function of the interaction strength $(\lambda/\omega_0)^2$ for $V/\Gamma=10$ and different $U$. The other parameters are as in Fig.~\ref{fig:DeltaIN_QuadLams_Vs}. The dotted lines indicate power laws $\sim(\lambda/\omega_0)^4$ for comparison, the dashed ones power laws $\sim (\lambda/\omega_0)^2$. Upper panel: bare data on a log-log scale. Lower panel: logarithmic derivative (apparent exponent).}
	\label{fig:DeltaIN_QuadLams_Us}
\end{figure}

\begin{figure}[t]
	\centering
	\includegraphics[width=1\linewidth]{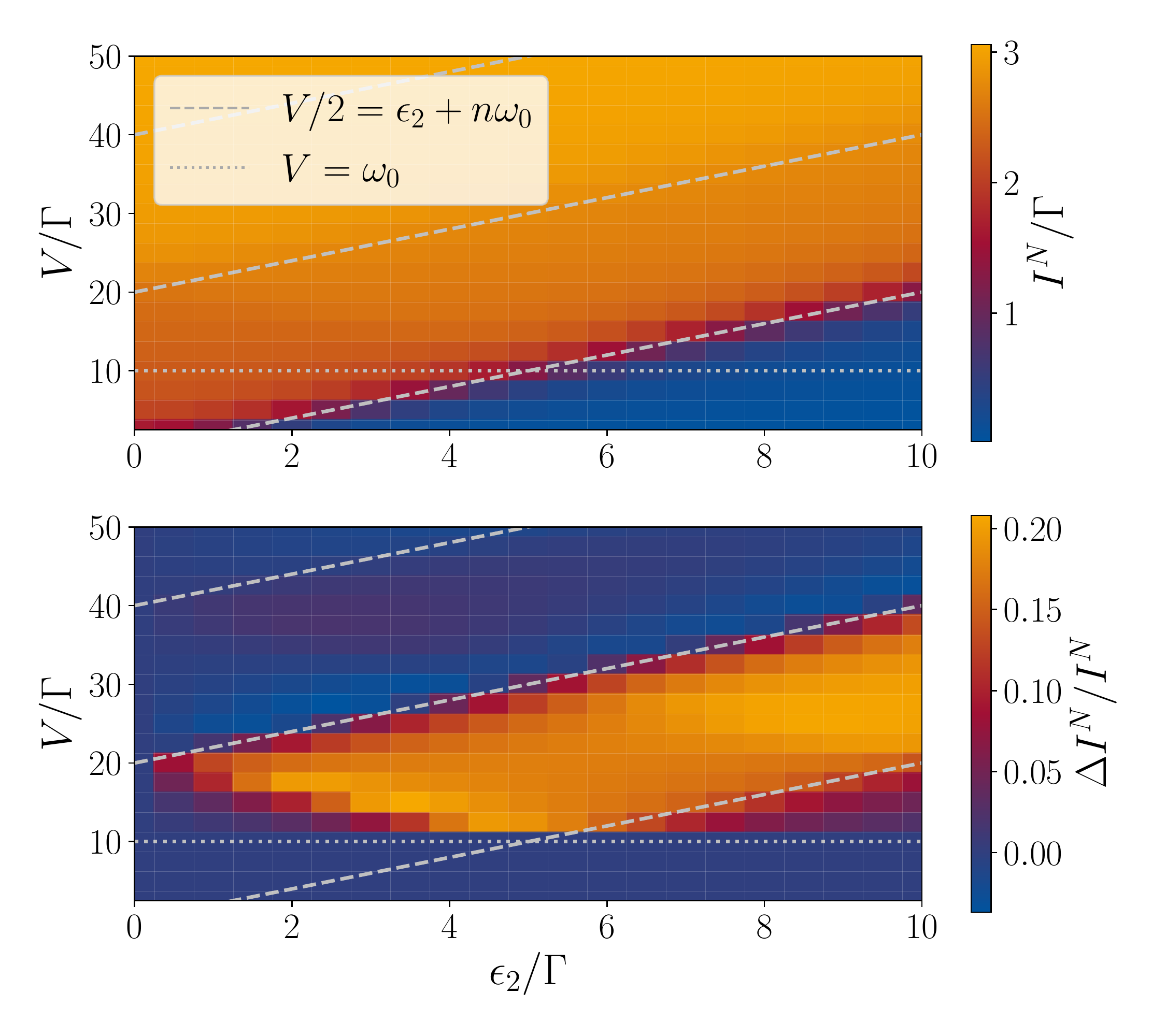}
	\caption{The particle current $I^{\rm N}$ (upper panel) and the relative violation of the current  conservation $\Delta I^{\rm N}/I^{\rm N}$ (lower panel) as functions of the on-site energy $\epsilon_2/\Gamma$ and voltage $V/\Gamma$ in the anti-adiabatic limit $\omega_0/\Gamma=10$. The temperature is $T_L=T_R=T_\text{p}=0$. We consider the SAHM limit $U=0$ for $\lambda/\omega_0=0.5$}
	\label{fig:DeltaIN_grid}
\end{figure}

However, by increasing the fermion-boson interaction strength $\lambda$ and by going deeper into the antiadiabatic regime $\Delta I^{\rm N}$ becomes sizable. This is exemplified in Fig.~\ref{fig:IN_largeDeltaIN}. In particular, for voltages around the phonon step, $\Delta I^{\rm N}$ becomes prominent. For $\lambda/\omega_0=1$ and $\omega_0/\Gamma =10$, $\Delta I^{\rm N}$ becomes so large that it even masks this step and the lowest order truncated functional RG results of $I^{\rm N}$ become useless for $2< V/\Gamma < 4$.

In the lowest-order truncation not all terms of order $U^2$, $\lambda^4$, and $U \lambda^2$ are kept. Thus $\Delta I^N$ must be of these orders. To investigate this, we first consider the SAHM limit $U=0$. In the upper panel of Fig.~\ref{fig:DeltaIN_QuadLams_Vs} we show $\Delta I^{\rm N}/I^{\rm N}$ (relative violation of the current conservation) as a function of $(\lambda/ \omega_0)^2$ for different voltages in the anti-adiabatic limit $\omega_0/\Gamma=10$. Note the log-log scale. For small $(\lambda/\omega_0)^2$ all curves follow a straight line of similar slope. This indicates power-law behavior in this regime. The dotted lines indicate power laws with exponent $2$. To further investigate this in the lower panel the logarithmic derivative of the data (apparent exponent; for a definition, see the $y$-axis label) is shown. If this becomes constant the bare data follow a power law with the constant being the exponent. For $\lambda \to 0$ all data sets approach the constant 2, as expected. Importantly, evaluating $\Delta I^{\rm N}$ provides a consistency check within the approximation scheme which indicates the reliability of the results. 

This analysis also indicates that our numerical solution of the RG flow equations is very accurate. In the logarithmic derivative any small numerical error in $I_{L/R}^{\rm N}$ would lead to a very large variation. Showing the consistency of the order of the breaking of conservation laws can thus also serve as a very useful tool to benchmark both the numerical implementation of the (nontrivial) RG flow equations as well as the numerical accuracy. This might become even more important in more complex models and for higher order truncation schemes, both leading to even more involved numerics \cite{Kloeckner2020}. 

We emphasize that by choosing the two-particle interaction sufficiently small the error due to the breaking of conservation laws can be made arbitrarily small. E.g., for the smallest $\lambda$ considered in Fig.~\ref{fig:DeltaIN_QuadLams_Vs} the relative error is less than $0.1\%$. However, in this limit also the interaction effects of interest might vanish. Compare, e.g., the phonon shoulders of Figs.~\ref{fig:IN_step_2} and \ref{fig:IN_largeDeltaIN}.

As an additional consistency check we show results for the relative breaking of current conservation as a function of $(\lambda/\omega_0)^2$ for different $U>0$ in Fig.~\ref{fig:DeltaIN_QuadLams_Us}. The other parameters are as in Fig.~\ref{fig:DeltaIN_QuadLams_Vs}. For very small $U/D$ the term ${\mathcal O} ([\lambda/\omega_0]^4)$ dominates at small $\lambda/\omega$. However, for larger $U/D$ we observe the expected crossover to a $\Delta I^{\rm N}/I^{\rm N} \sim (U/D) (\lambda/\omega_0)^2$ behavior.   

To obtain a better understanding of the dependencies of the particle current and of the relative breaking of current conservation on the parameters, in Fig.~\ref{fig:DeltaIN_grid} we show both as a function of the on-site energy $\epsilon_2$ and the voltage $V$ for $U=0$. As described in Sect.~\ref{sec:results}, the $I^{\rm N}(V)$ curve saturates for $V>\epsilon_2$ and shoulder-like features due to the phonon emission and absorption appear for $V/2\in[\epsilon_2+n \omega_0, \epsilon_2+(n+1) \omega_0]$, $n\in\mathbb{N}$. Unfortunately, this is also the regime, where the largest violation of the particle conservation can be observed. A sizeable $\Delta I^{\rm N}/I^{\rm N}$ is only found for $V>\omega_0$. 
Remind that $\Delta I^{\rm N}=0$ for $\epsilon_2=0$. Whereas the magnitude of the violation of current conservation increases gradually with increasing $\epsilon_2$, $\Delta I^{\rm N}$ seems to exhibit a step-like increase when $V$ passes the phonon frequency. 

\section{Summary}
\label{sec:summary}
We applied the functional RG on the Keldysh contour in a first order truncation to analyze a combination of the interacting resonant level model and the spinless Anderson-Holstein model, including both, a local (screened) Coulomb as well as a local fermion-phonon interaction in a single-level quantum dot region. Our focus was on the antiadiabatic regime as well as weak to intermediate interactions. 

The emergent low energy scale $T_{\rm K}$, characterizing the lead-dot tunneling is affected by both interaction terms. The coupling to the phonon mode leads to a reduction of this scale while the Coulomb interaction increases it. This manifests in the nonequilibrium steady-state particle current $I^{\rm N}$ through the quantum dot region, which we computed. The current-voltage characteristics show very rich behavior. It features a phononic step/shoulder as well as the power-law decay for large voltages earlier found for the IRLM. The characteristics of the two models superimpose in a straightforward way. We analyzed transport coefficients, in particular the electron contribution to the thermal conductance $\kappa_{\rm e}$. In both linear response as well as for finite bias voltages the effect of the Coulomb interaction can be fully included by expressing all energy variables in terms of $T_{\rm K}$ instead of $\Gamma$. An appropriate rescaling leads to a collapse of all curves for the same $\lambda$ but with different $U$ onto a single curve. 

We, finally, analyzed the violation of current conservation within first-order truncated functional RG. Depending on the model and physics to be investigated the nonconserving nature of the functional RG in its standard truncation can become one of the major obstacles in applying this approximation scheme. Our model and truncation order is ideally suited for such a case study as the inelastic fermion-phonon scattering leads to a broken current conservation while the numerics remains moderate and allows for a comprehensive investigation. The sum $\Delta I^{\rm N}$, of the left and right current can be used as a measure for the breaking of current conservation. It manifests to order $m_{\rm c}+1$ in the interactions $U$ and $\lambda^2$, with $m_{\rm c}$ being the truncation order ($m_{\rm c}=1$ in the present case). We showed that $\Delta I^{\rm N}$ indeed scales as the interactions squared. It thereby provides an internal consistency check for the reliability of the calculations in truncated functional RG.
Generally, we were able to confirm the existence of parameter regimes, where $\Delta I^{\rm N}$ is negligible but interesting nonequilibrium many-body physics is still observable, thus, consolidating the functional RG method as a useful tool. We, in particular, established that the violation of current conservation vanishes at particle-hole symmetry.

\section*{Acknowledgments}
We thank Andishe Khedri, Darvin Wanisch, and Severin Jakobs for discussions. This work was supported by the Deutsche Forschungsgemeinschaft (DFG, German Research Foundation) via RTG 1995 and under Germany’s Excellence Strategy-Cluster of Excellence Matter and Light for Quantum Computing (ML4Q) EXC2004/1 - 390534769. We acknowledge support from the Max Planck-New York City Center for Non-Equilibrium Quantum Phenomena. Simulations were performed with computing resources granted by RWTH Aachen University. 

\appendix*
\section{Numerical solution of the flow equations}
\label{app:frequency}

In the flow equations (\ref{eq:FlowEquation}) every self-energy component of the dot for a certain $\omega$ couples via the Greens function with a self-energy with other frequencies. Therefore, one obtains a set of coupled differential equations of first order. With this premise, the flow equations have to be solved on a finite and discrete frequency grid.
 
Because the grid needs to span a large frequency range compared to the energy-scale $\omega_0$, but also has to account for a strong frequency dependence close to $\epsilon_2$, a logarithmic grid symmetric around $\epsilon_2$ is used. During the flow, the center is shifted towards the renormalized static on-site energy $\Re\Sigma_{22}^\text{R}(\omega=0)$ successively.

In particular, for small $T$, where the steps introduced by the Fermi-functions become very sharp, the grid has to be sufficiently fine and the $\omega$ points have to be distributed as evenly as possible around this areas. Those are in general areas with a distance of order $\mathcal{O}(\omega_0)$ away from $\epsilon_2$. To this end, we use even more points in the interval $\omega\in\left[ - 4\omega_0, 4\omega_0 \right]$. This basic grid is then overlayed with further smaller logarithmic grids, which are symmetric around points showing characteristic and sharp features emerging at $n\omega_0 \pm V/2$ with $n\in \mathds{N}$. Such a logarithmic grid with $(N+1)$ points $\omega_k\in\left[-\omega_\text{max}, \omega_\text{max}\right]$ is constructed via
\begin{align}
	\omega_k = \omega_\text{max} \frac{2k-N}{N} \exp\left[ \frac{|N-2k|-N}{S} \right],
	k = 0,1,..,N.
\end{align}
In our calculations, $N$ is chosen to be odd, and therefore $\omega_k=0$ is excluded. Those grids are then shifted to be symmetric around the high resolution points.
The parameter $S$ is the resolution and is chosen in the following calculations such that $\min|\omega_k|=10^{- 3}{\Gamma}$. Different $N$ and $\omega_\text{max}$ were tested for various parameter combinations. For the basic grid, $\omega_\text{max}=10^{5}\Gamma$ and $N=6000$ was observed to result in converged results for the considered observables. For the finer grids $\omega_\text{max}=\Gamma$ and $N=200$ seems to be sufficient. In general, this results in grids of the size $N=7000$ to $N=11000$ points. The convergence of the calculations concerning this grid has been verified. 

A further difficulty arises as $\Sigma_{22}^x(\omega-s\omega_0)$, $x$ has to be evaluated for arbitrary frequencies along the flow. To evaluate the self-energy at frequencies between the grid-points, linear interpolation is used.
All frequency integrals are evaluated with the simple Riemann sum
\begin{align}
	\int\dd\omega F(\omega) \rightarrow \sum_k \frac{F(\omega_{k+1}) + F(\omega_{k})}{2} \left( \omega_{k+1} - \omega_{k} \right),
\end{align}
with $\omega_k$ being the grid points. This minimizes calculation time, in contrast, to, e.g., adaptive integration methods and was tested to give the same results for the used grid.
When calculating the Cauchy principle integral in Eq.~(\ref{eq:FlowEquation}), the grid has to be temporarily symmetrized around the pole of the integrand, where it is important not to incorporate the exact point of singularity. Because the numerator does not contain any singularities, we only have to consider the simple pole of the denominator. Due to those integrals, the self-energies are evaluated for every $\omega_k$ on the grid at every point of the flow. Therefore, the results at the end of the flow are very sensitive to the form of the grid, making the construction a highly non-trivial task.
The values of the self-energies for this temporary grid are also calculated via linear interpolation.

\end{document}